\RequirePackage{fix-cm}
\documentclass[twocolumn]{svjour3}          
\smartqed  
\usepackage{graphicx}
\usepackage[utf8]{inputenc}
\usepackage{float}
\usepackage{amsmath}
\usepackage{amssymb}
\usepackage{gensymb}
\usepackage{lmodern}
\usepackage{subcaption}
\usepackage{mwe}
\usepackage{tikz}
\usepackage{bbm}
\usepackage{listings}
\usepackage{xcolor}
\definecolor{codegreen}{rgb}{0,0.6,0}
\definecolor{codegray}{rgb}{0.5,0.5,0.5}
\definecolor{codepurple}{rgb}{0.58,0,0.82}
\definecolor{backcolour}{rgb}{0.95,0.95,0.92}

\lstdefinestyle{mystyle}{
  backgroundcolor=\color{backcolour},   commentstyle=\color{codegreen},
  keywordstyle=\color{magenta},
  numberstyle=\tiny\color{codegray},
  stringstyle=\color{codepurple},
  basicstyle=\ttfamily\footnotesize,
  breakatwhitespace=false,
  breaklines=true,
  captionpos=b,
  keepspaces=true,
  numbers=left,
  numbersep=5pt,
  showspaces=false,
  showstringspaces=false,
  showtabs=false,
  tabsize=2
}

\usepackage[margin=1in]{geometry}
\usepackage{mathtools}
\usepackage{graphicx}

\usepackage{comment}
\usepackage{enumerate}

\newcommand{\E}{\mathbb{E}}

\renewcommand{\phi}{\varphi}

\newcommand{\Var}{\mathrm{Var}}
\newcommand{\Cov}{\mathrm{Cov}}

\usepackage{graphicx}

\usepackage[numbers]{natbib}

\begin{document}

\title{The Development and Deployment of a Model for Hospital-level COVID-19 Associated Patient Demand Intervals from Consistent Estimators (DICE)
}


\author{Linying Yang \and Teng Zhang \and Peter Glynn \and David Scheinker  
}


\institute{Linying Yang \at
                Institute for Computational and Mathematical Engineering,
                Stanford University, Stanford, CA 94305, USA
            \and
            Teng Zhang \and
          Peter Glynn \and
          David Scheinker
          \at
              Department of Management Science and Engineering, Stanford University, Stanford, CA 94305, USA
}


\maketitle

\begin{abstract}
Hospitals commonly project demand for their services by combining their historical share of regional demand with forecasts of total regional demand. Hospital-specific forecasts of demand that provide prediction intervals, rather than point estimates, may facilitate better managerial decisions, especially when demand overage and underage are associated with high, asymmetric costs. Regional forecasts of patient demand are commonly available as a Poisson random variable, e.g., for the number of people requiring hospitalization due to an epidemic such as COVID-19. However, even in this common setting, no probabilistic, consistent, computationally tractable forecast is available for the fraction of patients in a region that a particular institution should expect. We introduce such a forecast, DICE (Demand Intervals from Consistent Estimators). We describe its development and deployment at an academic medical center in California during the `second wave' of COVID-19 in the Unite States. We show that DICE is consistent under mild assumptions and suitable for use with perfect, biased, unbiased regional forecasts. We evaluate its performance on empirical data from a large academic medical center as well as on synthetic data.
\end{abstract}

\section{Introduction}
\label{intro}

The COVID-19 pandemic has disrupted hospital operations the world over. Large influxes of patients requiring intensive care and mechanical ventilation have overwhelmed capacity, forced hospitals to triage, and have been associated with significantly elevated case fatality rates. Shortages of personal protective equipment (PPE) have exposed healthcare workers to additional risk and many have contracted COVID-19 and died.

Hospitals managers have a variety of options to increase total and available capacity when planning for an influx of COVID-19 patients \citep{weissman2020locally}. Managers may be able to increase total capacity by calling in additional nurses and doctors, opening previously closed beds, and acquiring additional PPE. Managers may be able to increase available capacity by expediting patient discharge or canceling or delaying non-discretionary, non-urgent patient admissions \citep{negopdiev2020elective}. The potential detriments to the quality of care and higher costs associated with these actions may be partially or fully mitigated if the decision to act is made with sufficient lead time. In the worst-case scenario when a hospital has insufficient intensive care unit (ICU) or ventilator capacity, patients with COVID-19 may experience significantly higher case mortality rates \cite{livingston2020coronavirus}. In less dire scenarios, nurses called in to work on short notice may require overtime pay while those scheduled a week in advance may not; PPE is less expensive when its purchase is not expedited; and patients whose non-urgent procedures are scheduled for later will experience less disruption than patients whose procedures are cancelled on short notice. In the United States, where healthcare is paid for through a combination of private and public insurance, the pandemic has created the additional challenge of significant financial stress as COVID-19 patients are associated with lower rates of reimbursement than patients who receive non-urgent, non-discretionary procedures such as tumor removal surgery or chemotherapy \cite{khullar2020covid}.

The complementary challenges of ensuring sufficient capacity to meet the demand associated with COVID-19 while avoiding unnecessarily long delays to non-COVID-19 care, require hospital managers to generate forecasts of the volume of COVID-19 patients requiring care at their institution. Managerial decisions based on forecasts of COVID-19 may benefit from the availability of the forecast with as much lead time as possible. To allow managers to account for the asymmetric risk associated with having insufficient capacity to meet urgent COVID-19 demand or non-urgent procedural demand, such forecasts should provide probabilistic, rather than point, estimates.

Our methodology reflects the random fluctuations that arise  at the hospital level that are averaged out at the regional level. For example, if a hospital receives, on average,  5\% of a county's hospitalization and the forecast county hospitalization level is 100, the random fluctuation about the mean hospital load of 5 patients can be significant (in a relative error sense). In particular our methodology provides a prediction interval on the number of COVID-19 positive patients at a given hospital rather than a ``point forecast''. In addition, our methodology takes into account the additional uncertainty induced by estimation error associated with estimating the underlying statistical parameters from observed data.

This paper is concerned with developing statistical methods to support hospital decision making with regard to COVID-19 capacity planning issues. In particular, hospital leadership can benefit from statistical tools to help them assess the amount of capacity that will need to be assigned to coronavirus patients in the weeks to come. A serious complication is that epidemiological forecasts typically focus on aggregate COVID-19 predictions that are provided at the regional level. For example, in California, the available COVID-19 forecasts are provided at the county level. Our goal in this paper is to provide a statistically principled methodology for obtaining hospital-level coronavirus hospitalization forecasts from such regional forecasts. A simple example is illustrative.

Consider a hospital preparing for a forecast influx of COVID-19 patients. Suppose the hospital can accommodate up to 20 COVID-19 patients with its current capacity, 21-30 COVID-19 patients by calling in additional staff, and over 30 patients by cancelling scheduled procedures. A forecast that the hospital will have to accommodate 25 .

Given a regional forecast for the daily number of hospitalizations as well as historical data on the share of regional hospitalizations accommodated by a specific hospital, all assumed to be Poisson random variables, we develop a forecast model DICE (Demand Intervals from Consistent Estimators). The primary contributions of this paper are as follows.
\begin{itemize}
    \item We show that DICE is consistent under mild assumptions and suitable for use with biased and unbiased regional forecasts.
    \item We show that DICE performed well on empirical data from a large academic medical center in California as well as on synthetic data.
    \item We describe the COVID-19 related capacity management decisions facilitated by the use of DICE.
\end{itemize}

\section{Literature}
Numerous COVID forecasting models have been developed since the start of the pandemic. Among them, many forecast regional-level COVID-19 cases, hospitalizations, and deaths \cite{covidactnow},
\cite{gleam}, \cite{jhulemaitre2020scenario}, \cite{berkeleyaltieri2020curating} and \cite{ferstad2020model}. Most such models use publicly available data and epidemic models to forecast hospitalizations down to the level of a single \textit{county} or several adjacent counties. However, few tools are available for \textit{hospitals} to make a probabilistic forecast of their expected share of the forecast regional volume. The data available to make such a forecast include: the outputs of the aforementioned models; detailed historical data on county-level hospitalizations, available from the national authorities such as \cite{cdc}; real-time data on hospitalizations in a particular county or region available from local authorities such as and \cite{calcat}; and hospital-specific hospitalization data to the managers of the institution generating the forecast.
\section{Setting}

This model was developed in response to a request from the COVID-19 planning leadership of a large academic medical center (AMC) in a large county in California during the summer of 2020. After the initial wave of COVID-19 cases was brought under control with non-pharmaceutical interventions such as social distancing, the hospital restarted non-urgent admissions for procedures such as surgery. As national news of a `second wave' of COVID-19 hospitalizations spread, the AMC leadership wanted to prepare. They requested a forecast that would inform them, with as much notice as possible, of an influx of COVID-19 patients sufficiently large that elective admissions should be halted in order to make capacity available for the expected COVID-19 patients. We were provided with the hospital's historical data on the number of admissions and the length of stay of each patient in the ACU and ICU, historical and forecast data for the total number of hospitalizations in the county, and automated daily updates on the number of new COVID-19 admissions to the ICU and ACU as well as patients currently in those units. We worked with hospital leadership to estimate the capacity of COVID-19 patients that the institution could accommodate without having to increase available capacity by canceling scheduled procedures as. We also worked with the leadership to determine an order for cancelling scheduled, non-urgent surgical procedures if necessary. The order was based primarily on the clinical acuity of those requiring the procedure, the average ICU and ACU post-operative length of stay associated with the procedure, and additional constraints on hospital operations. Those efforts are described in another work in progress. The goal of the present work was to generate a forecast that used recent data on the share of all COVID-19 patients in the county to provide two weeks notice for when the demand associated with COVID-19 patients was likely to require the cancellation of the scheduled procedures. Since hospital occupancy fluctuates naturally, rather than determine a hard cut-off for cancelling procedures, hospital leadership requested that we notify them if the upper bound of the prediction interval exceeded a pre-specified lower bound at which point they would evaluate the prospect of cancelling cases.

\section{Prediction Intervals with Perfect Forecasts}

We start by describing the problem setting from a mathematical perspective. We assume that we are currently in day $0$ and have been tasked with producing prediction intervals for the future number of hospital-level ACUte care unit (ACU) and intensive care unit (ICU) hospitalization at the end of day $r$, with $r\geq 0$. For the purpose of predicting these hospital-level prediction intervals, we have available historical data $((A_j,B_j,N_j): -n\leq j\leq -1)$, where $N_j$ is the total number of regional hospitalizations at the end of day $j$, $A_j$ is the number of ACUte care hospitalizations at the given hospital at the conclusion of day $j$, and $B_j$ is the number of ICU hospitalizations at the given hospital at the end of day $j$. Furthermore, we assume that we have available a point forecast $F_r$ for the mean number of regional hospitalizations at the end of day $r$.









Throughout the paper, we take the view that the $A_j$'s,$B_j$'s and $N_j$'s can be reasonably modeled as Poisson distributed random variables (rv's).We will use the notation $\mathcal{P}(\lambda)$ to denote a Poisson rv with mean $\lambda$. There is an extensive mathematical theory supporting the use of Poisson rv's in the setting of such count statistics; see, for example, \cite{daley2003introduction}.

A simple model relates $A_j$ and $B_j$ to $N_j$ by assuming that $\E A_j=p_0\E N_j$ and $\E B_j = q_0 \E N_j$ for $p_0,q_0 \geq 0$. Because the $N_j$'s are subject to episodic epidemic growth spurts, \textit{we do not assume that $\E N_j$ is constant.} Instead, we permit $\lambda_j \equiv \E N_j$ to fluctuate in a potentially complex fashion.

In this section, we assume that point forecast $F_r$ is $\textit{perfect}$, in the sense that
\begin{align}\label{eq:forecast-perfect-assume}
    F_r=\lambda_r. \tag{2.1}
\end{align}
It follows that if we select $l(\lambda)$ (the lower endpoint) as the largest integer such that $P(\mathcal{P}(\lambda)<l(\lambda))\leq \frac{\delta}{2}$ and  $u(\lambda)$ (the upper endpoint) as the smallest integer such that $P(\mathcal{P}(\lambda)>u(\lambda))\leq \frac{\delta}{2}$, then $[l(F_r),u(F_r)]$ is a $100(1-\delta)\%$ \textit{prediction interval} for $N_r$ having the property that $P(N_r\in[l(F_r),u(F_r)])\geq 1-\delta$.

To obtain similar prediction intervals for $A_r$ and $B_r$, we need to estimate $p_0$ and $q_0$ from the data. The obvious estimators for $p_0$ and $q_0$ are given by
\begin{equation}\label{eq:estimation-hat-p-hat-q}
\begin{aligned}
    \hat{p} &= \sum_{j=-n}^{-1} A_j/\sum_{j=-n}^{-1} N_j,\\
    \hat{q} &= \sum_{j=-n}^{-1} B_j/\sum_{j=-n}^{-1} N_j.
\end{aligned} \tag{2.2}
\end{equation}

In fact, $\hat p$ and $\hat q$ are the maximum likelihood estimators (MLE's) for $p_0$ and $q_0$ when $A_j$ (and $B_j$) are, conditional on $N_{-n}, \dots, N_{-1}$, independent in $j$ and binomially distributed with parameters $N_j$ and $p_0$ (and $q_0$).

This leads to the prediction intervals $[l(\hat{p}F_r), \\u(\hat{p}F_r)]$ for $A_r$ and $[l(\hat{q}F_r),u(\hat{q}F_r)]$ for $B_r$. We refer to these prediction intervals as the \textit{plug-in prediction intervals} based on perfect forecasts.

\section{Prediction Intervals with Perfect Forecasts: Incorporating Estimation Uncertainty}

Our frequentist approach starts by setting $\theta = (p, q)$ and letting $P_\theta(\cdot)$ be the probability model under which the $(A_i, B_i, N_i-A_i-B_i)$'s are conditionally independent given the $N_j$'s., with $(A_i, B_i, N_i-A_i-B_i)$ following a multinomial distribution with parameters $(N_i, p_0, q_0, 1-p_0-q_0)$. Our ideal prediction interval for $A_r$ would, of course, be the interval $[\ell(p_0F_r), u(p_0F_r)]$.
Since $p_0$ is unknown, the plug-in interval $[\ell(\hat pF_r), u(\hat pF_r)]$ of section 2 is an obvious alternative.
However, because $\hat p$ is random, we can not guarantee that
\begin{align}\label{eq:prob-bound-perfect-uncertainty}
    P_{\theta_0} (A_r < \ell(\hat  p F_r ) ) \le \delta /2, \tag{3.1}
\end{align}
where $\theta_0 = (p_0, q_0)$.

Instead, we seek a probabilistic guarantee, namely that~\eqref{eq:prob-bound-perfect-uncertainty} holds, with probability (or confidence level) $1 - \alpha$.

We can accomplish this by choosing the integer $z_\ell$ so that
\begin{align}\label{eq:prob-bound-perfect-uncertainty-z_ell}
    P_{\theta_0} \left( \ell(\hat p F_r) - \ell(p_0F_r) \le z_\ell  \right) \ge 1-\alpha. \tag{3.2}
\end{align}
On the event $\{ \ell(p_0 F_r) \ge  \ell(\hat p F_r) - z_\ell  \}$,
\begin{align*}
    P_{\theta_0} \left( A_r <  \ell(\hat pF_r) - z_\ell   \right)
    \le
    P_{\theta_0} \left( A_r < \ell( p_0F_r) \right)
    \le \delta / 2.
\end{align*}
Hence, with confidence at least $1 - \alpha$, $ \ell(\hat pF_r) - z_\ell $ is an appropriately chosen value for the left endpoint of $A_r$'s prediction interval.

Similarly, if we choose the integer $z_r$ so that
\begin{align}\label{eq:prob-bound-perfect-uncertainty-z_r}
    P_{\theta_0} \left( u(\hat p F_r) - u(p_0F_r) \ge z_r  \right) \ge 1-\alpha, \tag{3.3}
\end{align}
$ u(\hat pF_r) - z_r $ is a right endpoint for which\\ $ P_{\theta_0} \left( A_r > u(\hat pF_r) - z_r   \right) \le \delta /2 $ holds, at a confidence level of at least $1-\alpha$.
Hence, we adopt the interval $[ \ell(\hat pF_r) - z_\ell ,  u(\hat pF_r) - z_r ]$ as our prediction interval for $A_r$ that takes into account the estimation uncertainty that is present in $\hat p$.

To compute $z_\ell$ and $z_r$ from~\eqref{eq:prob-bound-perfect-uncertainty-z_ell} and~\eqref{eq:prob-bound-perfect-uncertainty-z_r}, we use the parametric bootstrap (see, for example, \cite{efron1994introduction}), thereby computing the values $z^*_\ell$ and $z^*_r$ such that
\begin{align*}
    P_{\hat \theta} \left( \ell(\hat p^* F_r) - \ell(\hat pF_r) \le z^*_\ell  \right) \ge 1-\alpha
\end{align*}
and
\begin{align*}
    P_{\hat \theta} \left( u(\hat p^* F_r) - u(\hat pF_r) \ge z^*_r  \right) \ge 1-\alpha,
\end{align*}
where $\hat \theta = (\hat p, \hat q)$ and $\hat p^*$ is the estimator for $\hat p$ obtained from a bootstrap sample of the data set;
the details can be found in the algorithm as described below. This leads to the prediction interval $[ \ell(\hat pF_r) - z^*_\ell  ,  u(\hat pF_r) - z^*_r ]$;
we refer to this as the \textit{bootstrap} prediction interval for $A_r$ based on perfect forecasts.
We can similarly compute the bootstrap prediction interval for $B_r$ based on perfect predictions.

Specifically, our bootstrap prediction intervals are produced by the following algorithm.
\newline

\smallskip

\noindent
\textit{Algorithm 1}.

\begin{enumerate}[1.]

\item Simulate independent Poisson random variables $(N_i^*: -n \le i \le -1)$ with mean $(F_i: -n\le i \le -1)$.

\item Conditional on $N_i^*$, simulate a multinomial rv $(A_i^*, B_i^*, N_i^*-A_i^*-B_i^*)$ with parameters $(n, \hat p, \hat q,\\ 1-\hat p- \hat q)$.

\item Compute
\begin{align*}
    \hat{p}^*&=\frac{\sum_{j=-n}^{-1}A_j^*}{\sum_{j=-n}^{-1}N_j^*}\\
    \hat{q}^*&=\frac{\sum_{j=-n}^{-1}B_j^*}{\sum_{j=-n}^{-1}N_j^*}
\end{align*}

\item Compute $(\ell (\hat p^*F_r), u (\hat p^*F_r), \ell (\hat q^*F_r), u (\hat q^*F_r) )$.

\item Repeat steps 1 to 4 $b$ times, thereby yielding $b$ 4-tuples $(\ell (\hat p_i^*F_r), u (\hat p_i^*F_r), \ell (\hat q_i^*F_r), u (\hat q_i^*F_r) )$.

\item Compute the smallest integers $z^*_{A, \ell}$ and $z^*_{B, \ell} $ for which
\begin{align*}
    \frac{1}{b} \sum_{i = 1}^b I \left( \ell(\hat p_i^* F_r) - \ell(\hat pF_r) \le z^*_{A, \ell}  \right) \ge 1-\alpha
\end{align*}
and
\begin{align*}
    \frac{1}{b} \sum_{i = 1}^b I \left( \ell(\hat q_i^* F_r) - \ell(\hat qF_r) \le z^*_{B, \ell}  \right) \ge 1-\alpha,
\end{align*}
and the largest integers $z^*_{A, r}$ and $z^*_{B, r} $ for which
\begin{align*}
    \frac{1}{b} \sum_{i = 1}^b I \left( u(\hat p_i^* F_r) - u(\hat pF_r) \ge z^*_{A, r}  \right) \ge 1-\alpha
\end{align*}
and
\begin{align*}
    \frac{1}{b} \sum_{i = 1}^b I \left( u(\hat q_i^* F_r) - u(\hat qF_r) \ge z^*_{B, r}  \right) \ge 1-\alpha.
\end{align*}
Then, the intervals
$[ [\ell(\hat pF_r) - z^*_{A, \ell}]^+  ,   u(\hat pF_r) - z^*_{A, r} ]$
and
$[ [\ell(\hat qF_r) - z^*_{B, \ell}]^+  ,   u(\hat qF_r) - z^*_{B, r} ]$ are the bootstrap prediction intervals for $A_r$ and $B_r$ respectively, where $[x]^+  \stackrel{\Delta}{=} \max (x, 0)$ for $x \in \mathbb R$.
\end{enumerate}

\section{Unbiased Forecasts with Lognormal Errors}

The model described in Section 2 and 3 assumes no forecast error. As a consequence, the distribution for $N_i$ is Poisson distributed with mean $F_i$. However, the forecast $F_i$ itself is imperfect, and there typically is additional uncertainty in the prediction of $N_i$ (beyond the stochastic variability of a Poisson rv) that should be reflected in the prediction interval.
In this section, we model the forecast error by assuming that
\begin{align} \label{eq:assumption-forecast-error-gamma}
    F_i = \lambda_i \Gamma_i^{-1}, \tag{4.1}
\end{align}
where $N_i$ is (again) Poisson with mean $\lambda_i$, and the relative forecast error $\Gamma_i^{-1}$ is assumed to be log-normally distributed.
Furthermore, we assume that the $N_i$'s are independent of the $\Gamma_i^{-1}$'s, and that the forecasts are \textit{relatively unbiased}, in the sense that
\begin{align} \label{eq:assumption-relative-unbiased}
    \E\left( \frac{N_i}{F_i} \right) = 1 \tag{4.2}
\end{align}
for all $i$, thereby implying that $\E[ \Gamma_i ] = 1$.

Of course, one expects that if the forecast under-predicts $N_i$ at time $i$, $F_{i+1}$ is also likely to under-predict $N_{i+1}$. This suggests that the $\Gamma_i$'s should be modeled as a correlated sequence.
In particular, we will assume that if $Y_i = \log \Gamma_i$, the $Y_i$'s form a stationary sequence that evolves according to the recursion
\begin{align*}
    Y_{i+1} = \rho_0 Y_i + Z_{i+1},
\end{align*}
where the $Z_i$'s are independent and identically distributed (iid) normally distributed rv's with mean $\mu_0$ and variance $\sigma^2_0$.
Note that the stationarity of the $Y_i$'s implies that $\rho \in (-1, 1)$, with $Y_i$ having a normal distribution having mean $\mu_0 (1 - \rho_0)^{-1}$ and variance $\sigma_0^2(1-\rho_0^2)^{-1}$; see \cite{anderson1971statistical}.

For this model, we need to estimate the parameters $\mu_0, \sigma_0^2$ and $\rho_0$ associated with the log-normally distributed forecast error sequence. As in Section 2 and 3, we assume that we have observed the time series $((A_i, B_i, A_i, F_i ): -n \le i \le -1)$, and we adopt the view that we wish to impose as few assumptions as possible on the $\lambda_i$'s (given the episodic nature of the coronavirus epidemic). For this reason, we will use the method of moments to estimate $\mu_0, \sigma_0^2$ and $\rho_0$.

Given~\eqref{eq:assumption-relative-unbiased}, we require that
\begin{align*}
    \E [\exp (Y_i)]
    &= \exp\left( \frac{\mu_0}{1 - \rho_0} +\frac{1}{2} \frac{\sigma_0^2}{1-\rho_0^2}  \right)\\
    &\stackrel{\Delta}{=}
    m_1(\mu_0, \sigma_0^2, \rho_0)
    =
    1.
\end{align*}
To obtain a second equation, note that
\begin{align*}
    \E \left[ \frac{N_i^2 - N_i}{F_i^2} \right]
    & = E[N_i^2 - N_i]\cdot \lambda_i^{-2} E[\Gamma_i^{2}] \\
    & = ( (\lambda_i + \lambda_i^2) - \lambda_i ) \cdot\lambda_i^{-2} E[\Gamma_i^{2}] \\
    & = E[ \exp (2Y_i)] \\
    & = \exp\left( \frac{2\mu_0}{1 - \rho_0} +2 \frac{\sigma_0^2}{1-\rho_0^2}  \right) \\
    &\stackrel{\Delta}{=}
    m_2(\mu_0, \sigma_0^2, \rho_0).
\end{align*}
For the third equation, we observe that
\begin{align*}
    \E \left[ \frac{N_i N_{i+1}}{F_i F_{i+1}} \right]
    & = \E N_i \cdot \E N_{i+1}
    \cdot (\lambda_i \lambda_{i+1})^{-1} \E \Gamma_i \Gamma_{i+1} \\
    & = \E[\exp(Y_i + Y_{i+1})] \\
    & = \E [\exp((1+\rho_0)Y_i + Z_{i+1})] \\
    & = \exp\left( \frac{2\mu_0}{1 - \rho_0} + \frac{\sigma_0^2}{1-\rho_0}  \right)\\
    &\stackrel{\Delta}{=}
    m_3(\mu_0, \sigma_0^2, \rho_0)
\end{align*}
This suggests that we estimate $\mu_0, \sigma_0^2$ and $\rho_0$ by minimizing the objective
\begin{align*}
    \left(\hat M_2 - m_2(\mu, \sigma^2, \rho)\right)^2 +
    \left(\hat M_3 - m_3(\mu, \sigma^2, \rho)\right)^2
\end{align*}
subject to
\begin{align*}
    m_1(\mu, \sigma^2, \rho) = 1, \\
    -1 \le \rho \le 1, \\
    \sigma^2 \ge 0,
\end{align*}
where
\begin{align*}
    \hat M_2
    & = \frac{1}{n} \sum_{i = -n}^{-1} \left( \frac{N_i^2 - N_i}{F_i^2}, \right),\\
    \hat M_3
    & = \frac{1}{n-1} \sum_{i = -n+1}^{-1} \left( \frac{N_iN_{i-1}}{F_iF_{i-1}} \right),
\end{align*}
followed by utilizing the minimizer $(\hat \mu, \hat \sigma^2, \hat \rho)$ as our estimator of $(\mu_0, \sigma_0^2, \rho_0)$, and then estimate $\hat p$ and $\hat q$ as in~\eqref{eq:estimation-hat-p-hat-q}.
When $n$ is large (and the statistical model describes the data well), we expect that the objective function will vanish at $(\hat \mu, \hat \sigma^2, \hat \rho)$, in which case
\begin{align*}
    m_i(\hat \mu, \hat \sigma^2, \hat \rho) = \hat M_i
\end{align*}
will be satisfied as equations for $i = 2, 3$. In the Appendix, we prove that our estimators for $\mu_0, \sigma_0^2$, and $\rho_0$ are consistent, under very moderate assumptions on the $\lambda_i$'s.

We note that in this model, the prediction interval for $N_r$ must reflect the additional randomness stemming from the fact that the mean of the Poisson random variable is itself random, namely it is given by $F_r \Gamma_r$.
In particular, let $\mathcal P (\mu, \sigma^2, \rho, f)$ be a rv that is conditionally Poisson distributed, with (random) mean $f\exp(N(\mu/(1-\rho), \sigma^2/(1-\rho^2))) $, where $N(\mu/(1-\rho), \sigma^2/(1-\rho^2))$ is a normal rv with mean $\mu/(1-\rho)$ and variance $\sigma^2/(1-\rho^2)$.
The plug-in prediction interval for $A_r$ based on this model is the interval $ [ \ell (\hat \mu, \hat \sigma^2, \hat \rho, \hat p F_r), u(\hat \mu, \hat \sigma^2, \hat \rho, \hat p F_r)  )]$, where $\ell ( \mu,  \sigma^2,  \rho, f)$ is the largest integer $j$ such that\\$P( \mathcal P (\mu, \sigma^2, \rho, f) < j ) \le \delta /2$ and $u( \mu,  \sigma^2,  \rho, f) $ is the smallest integer $k$ such that
$P( \mathcal P (\mu, \sigma^2, \rho, f) > k ) \le \delta /2$.
Similarly, $[ \ell (\hat \mu, \hat \sigma^2, \hat \rho, \hat q F_r), u(\hat \mu, \hat \sigma^2, \hat \rho, \hat q F_r)  )]$ is the plug-in prediction interval for $B_r$.

The computation of $ \ell ( \mu,  \sigma^2,  \rho, f) $ and $u ( \mu,  \sigma^2,  \rho, f)$ can be implemented via Monte Carlo, using the following algorithm.

\bigskip

\noindent
\textit{Algorithm 2}

\begin{enumerate} [1.]
    \item Simulate $Y_r$ as a normal rv with mean $\mu / (1-\rho)$ and variance $\sigma^2/(1-\rho^2)$.

    \item Generate $N_r$ as a Poisson rv with mean $f \exp (Y_r)$.

    \item Repeat Steps 1 and 2, independently, $m$ times, thereby yielding $N_{r,1}, \dots, N_{r,m}$.

    \item Define the estimator $\hat \ell (\mu, \sigma^2, \rho, f)$ for $\ell (\mu, \sigma^2, \rho, f)$ as the largest integer $j$ such that
    \begin{align*}
        \frac{1}{m}
        \sum_{i = 1}^m I(N_{r, i} < j) \le \delta/2
    \end{align*}
    and define the estimator $\hat u (\mu, \sigma^2, \rho, f)$ for\\$u (\mu, \sigma^2, \rho, f)$ as the smallest integer $k$ for which
    \begin{align*}
        \frac{1}{m}
        \sum_{i = 1}^m I(N_{r, i} > k) \le \delta/2.
    \end{align*}
\end{enumerate}
We now turn to the construction of prediction intervals for $A_r$ and $B_r$ that reflect the additional uncertainty due to the need to estimate $\mu_0, \sigma^2_0, \rho_0$ fro the observed data $((A_i, B_i, N_i, F_i): -n \le i \le -1)$.
Again, we use the bootstrap to compute the corrections $z^*_{A, \ell}, z^*_{B, \ell}, z^*_{A, r}, z^*_{B, r}$ that appear in this setting (that are direct analogs to those appearing in Algorithm 1 for perfect forecasts.)

\bigskip

\noindent
\textit{Algorithm 3}

\begin{enumerate} [1.]
    \item Generate $Y^*_{-n}$ as a normal rv with mean $\hat \mu /(1-\hat \rho)$ and variance $\hat \sigma^2 / (1 - \hat \rho^2)$.

    \item For $-n < i \le -1$, simulate $Y_i^*$ via the recursion
    \begin{align*}
        Y_i^* = \hat \rho Y_{i-1}^* + Z_i^*,
    \end{align*}
    where the $Z_i^*$'s are independently simulated as normal rv's with mean $\hat \mu$ and variance $\hat \sigma^2$.

    \item Given $(Y_i^*: -n \le i \le -1)$, simulate the $N_i^*$'s as independent Poisson rv's with means $( F_i\exp(Y_i^*): -n\le i \le -1 )$.

    \item Compute
    \begin{align*}
        \hat M_2^*
        &=
        \frac{1}{n} \sum_{i = -n}^{-1} \left( \frac{N_i^{*2} - N_i^*}{F_i^2} \right), \\
        \hat M_3^*
        &=
        \frac{1}{n-1} \sum_{i = -n+1}^{-1} \left( \frac{N_i^{*}N_{i-1}^*}{F_i F_{i-1}} \right).
    \end{align*}

    \item Compute the minimizer $(\hat{\mu}^*, \hat{\sigma}^{2*}, \hat{\rho}^*)$ of
    \begin{align*}
        \left(\hat M_2^* - m_2(\mu, \sigma^2, \rho)\right)^2 +
    \left(\hat M_3^* - m_3(\mu, \sigma^2, \rho)\right)^2
    \end{align*}
    subject to
    \begin{align*}
        m_1(\mu, \sigma^2, \rho) = 1, \\
        -1 \le \rho \le 1, \\
        \sigma^2 \ge 0.
    \end{align*}

    \item Generate $(A_i^*, B_i^*, N_i^* - A_i^* -B_i^*)$ as multinomial rv's with parameters $(N_i^*, \hat p, \hat q, 1 - \hat p - \hat q)$, $-n \le i \le -1$.

    \item Compute
    \begin{align*}
    \hat{p}^*&=\frac{\sum_{j=-n}^{-1}A_j^*}{\sum_{j=-n}^{-1}N_j^*}\\
    \hat{q}^*&=\frac{\sum_{j=-n}^{-1}B_j^*}{\sum_{j=-n}^{-1}N_j^*}
    \end{align*}

    \item Use Algorithm 2 to compute $\ell (\hat \mu^*, \hat \sigma^{2*}, \hat \rho^*, \hat p^* F_r), \\
    u (\hat \mu^*, \hat \sigma^{2*}, \hat \rho^*, \hat p^* F_r), \ell (\hat \mu^*, \hat \sigma^{2*}, \hat \rho^*, \hat q^* F_r), \\
    u (\hat \mu^*, \hat \sigma^{2*}, \hat \rho^*, \hat q^* F_r)$.

    \item Repeat Steps 1 to 8 $b$ times, thereby yielding $b$ 4-tuples
    $(
    \ell (\hat \mu_i^*, \hat \sigma_i^{2*}, \hat \rho_i^*, \hat p_i^* F_r),
    u (\hat \mu_i^*, \hat \sigma_i^{2*}, \hat \rho_i^*, \hat p_i^* F_r), \\
    \ell (\hat \mu_i^*, \hat \sigma_i^{2*}, \hat \rho_i^*, \hat q_i^* F_r),
    u (\hat \mu_i^*, \hat \sigma_i^{2*}, \hat \rho_i^*, \hat q_i^* F_r)
    )$.

    \item Compute the smallest integers $z_{A, \ell}^*$ and $z_{B, \ell}^*$ for which
    \begin{align*}
        \frac{1}{b} \sum_{i = 1}^b I ( & \ell(\hat \mu_i^*, \hat \sigma_i^{2*}, \hat \rho_i^*, \hat p_i^* F_r) - \\
        & \ell(\hat \mu, \hat \sigma^{2}, \hat \rho, \hat p F_r)
         \le z^*_{A, \ell} ) \ge 1-\alpha
    \end{align*}
    and
    \begin{align*}
        \frac{1}{b} \sum_{i = 1}^b I  (& \ell(\hat \mu_i^*, \hat \sigma_i^{2*}, \hat \rho_i^*, \hat q_i F_r) - \\
        &\ell(\hat \mu, \hat \sigma^{2}, \hat \rho, \hat q F_r) \le z^*_{B, \ell}  ) \ge 1-\alpha
    \end{align*}
    and the largest integers $z_{A, r}^*, z_{B, r}^*$ for which
    \begin{align*}
        \frac{1}{b} \sum_{i = 1}^b I ( & \ell(\hat \mu_i^*, \hat \sigma_i^{2*}, \hat \rho_i^*, \hat p_i^* F_r) - \\
        & \ell(\hat \mu, \hat \sigma^{2}, \hat \rho, \hat p F_r) \ge z^*_{A, r}  ) \ge 1-\alpha
    \end{align*}
    and
    \begin{align*}
        \frac{1}{b} \sum_{i = 1}^b I ( & \ell(\hat \mu_i^*, \hat \sigma_i^{2*}, \hat \rho_i^*, \hat q_i^* F_r) - \\
        & \ell(\hat \mu, \hat \sigma^{2}, \hat \rho, \hat q F_r) \ge z^*_{B, r}  ) \ge 1-\alpha
    \end{align*}
\end{enumerate}

Then, $ [ [\ell(\hat \mu, \hat \sigma^2, \hat \rho, \hat p F_r) - z^*_{A, \ell}]^+ ,
u(\hat \mu, \hat \sigma^2, \hat \rho, \hat p F_r) - z^*_{A, r}] $ and
$ [ [\ell(\hat \mu, \hat \sigma^2, \hat \rho, \hat q F_r) - z^*_{B, \ell}]^+ ,
u(\hat \mu, \hat \sigma^2, \hat \rho, \hat q F_r) - z^*_{B, r}] $
are our bootstrap prediction intervals for $A_r$ and $B_r$, respectively, based on unbiased log-normal forecasts.

\section{Biased Forecasts with Log-normal Errors}

We now modify the model of Section 4 to permit biased forecasts. The only change we make here is that we drop the requirement~\eqref{eq:assumption-relative-unbiased}. In this case, we need to add an additional moment identity in order to uniquely identify the coefficients $(\mu_0, \sigma^2_0, \rho_0)$ underlying the forecast errors given by the $\Gamma_i^{-1}$'s.
Note that
\begin{align*}
    \E\left( \frac{N_i}{F_i} \right)
    =
    \E N_i (\lambda_i^{-1}  \E \Gamma_i) = \E \Gamma_i
    = m_1 (\mu_0, \sigma^2_0, \rho_0).
\end{align*}
This suggests that we should estimate $(\mu_0, \sigma^2_0, \rho_0)$ via the minimizer $(\hat \mu, \hat \sigma^2, \hat \rho)$ of the objective function
\begin{align}\label{eq:minimizing-objective-sum-square-loss-3-moments}
    \sum_{i = 1}^3 \left(
    \hat M_i - m_i(\mu, \sigma^2, \rho)
    \right)^2 \tag{5.1}
\end{align}
subject to
\begin{align*}
    -1 \le \rho \le 1,\\
    \sigma^2 \ge 0,
\end{align*}
where $\hat M_2$ and $\hat M_3$ are defined as in Section 4 and
\begin{align*}
    \hat M_1 = \sum_{i = -n}^{-1} \left( \frac{N_i}{F_t} \right).
\end{align*}
As in Section 4, we estimate $p_0$ and $q_0$ via $\hat p$ and $\hat q$ as in~\eqref{eq:estimation-hat-p-hat-q}. As in Section 4, $ [ \ell (\hat \mu, \hat \sigma^2, \hat \rho, \hat p F_r), u(\hat \mu, \hat \sigma^2, \\ \hat \rho, \hat p F_r)  ]$ and $ [ \ell (\hat \mu, \hat \sigma^2, \hat \rho, \hat q F_r), u(\hat \mu, \hat \sigma^2, \hat \rho, \hat q F_r) ]$ are then our plug-in prediction intervals for $N_r$ based on the biased log-normal forecast error model.

Similarly, incorporating the estimation error related to estimating $(\mu_0, \sigma^2_0, \rho_0, p_0, q_0)$ via $(\hat \mu, \hat \sigma^2,\hat \rho, \\ \hat p, \hat q)$ requires only small modifications to the methodology of Section 4. The modified version of Algorithm 3 reflecting use of biased forecasts is provided next.

\bigskip

\noindent
\textit{Algorithm 4}

\smallskip

\noindent
Algorithm 4 is identical to Algorithm 3, excepting that $(\hat \mu, \hat \sigma^2,\hat \rho)$ is now the minimizer of~\eqref{eq:minimizing-objective-sum-square-loss-3-moments}, and Steps 4 and 5 are modified as follows:
\begin{enumerate}
    \item [$4^\prime.$] Compute
    \begin{align*}
        \hat M_1^*
        &=
        \frac{1}{n} \sum_{i = -n}^{-1} \left( \frac{N_i^*}{F_i} \right),\\
        \hat M_2^*
        &=
        \frac{1}{n} \sum_{i = -n}^{-1} \left( \frac{N_i^{*2} - N_i^*}{F_i^2} \right), \\
        \hat M_3^*
        &=
        \frac{1}{n-1} \sum_{i = -n+1}^{-1} \left( \frac{N_i^{*}N_{i-1}^*}{F_i F_{i-1}} \right).
    \end{align*}

    \item [$5^\prime.$]  Compute the minimizer $(\hat \mu^*, \hat \sigma^{2*}, \hat \rho^*)$ of
    \begin{align*}
        \sum_{i = 1}^3 \left(
        \hat M_i - m_i(\mu, \sigma^2, \rho)
        \right)^2
    \end{align*}
    subject to
    \begin{align*}
        -1 \le \rho \le 1, \\
        \sigma^2 \ge 0.
    \end{align*}
\end{enumerate}

Algorithm 4 yields our desired bootstrap prediction intervals for $A_r$ and $B_r$, just as Algorithm 3 yields such intervals for the unbiased model of Section 4.

\section{Empirical Results and Results on Synthetic Data}


\subsection{Model Deployment and Evaluation: Empirical Data}
\begin{figure*}
        \includegraphics[width=1\textwidth]{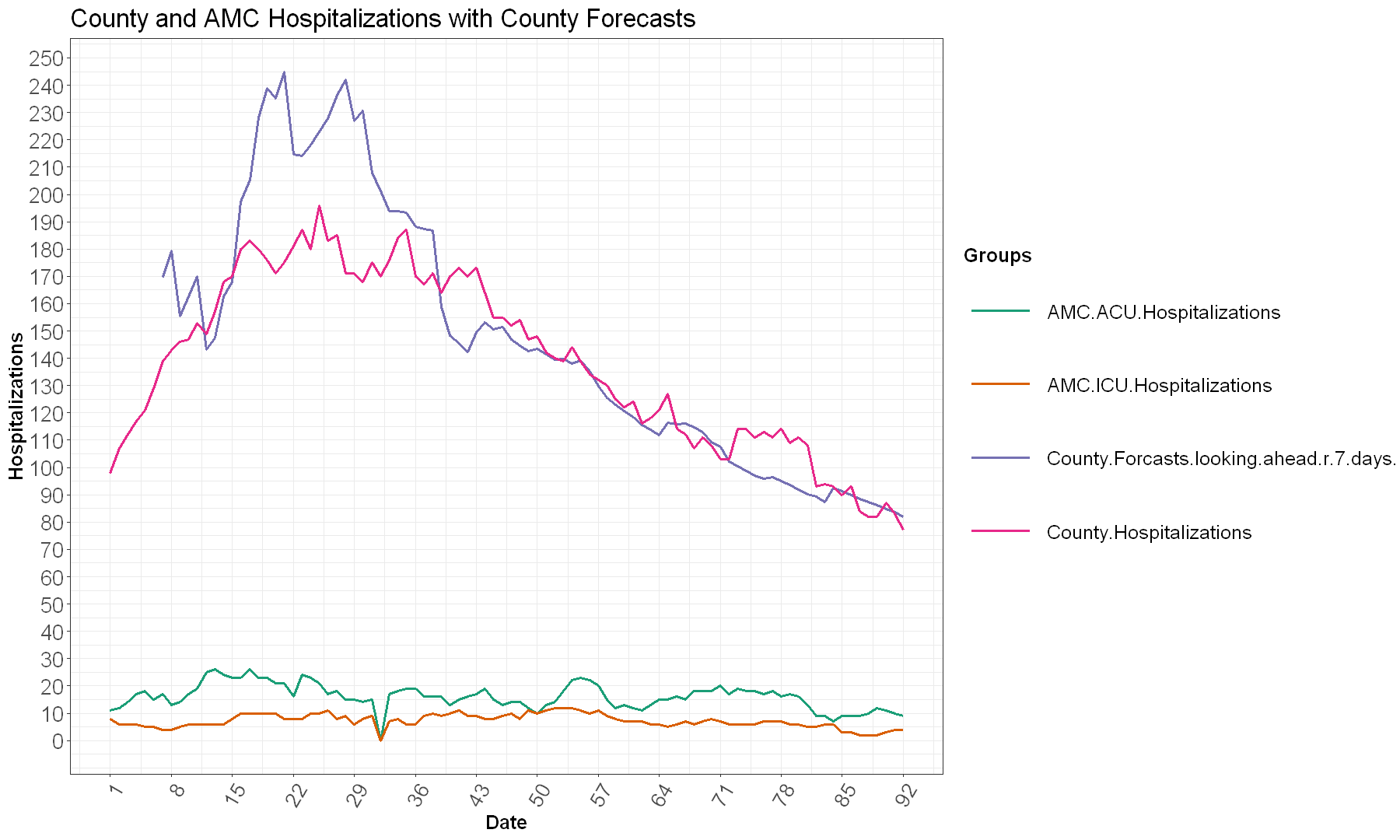}
        \caption{Empirical Data}
        \label{fig:lemma-predictions}
    \end{figure*}

We use historical county-level COVID-19 hospitalization forecasts and ACU, ICU COVID-19 hospitalizations from the AMC studied. Given the small number of patients, we protect patient privacy by replacing the actual date with the number of days from a reference date during the summer of 2020. The values are shown in Figure~\ref{fig:lemma-predictions}.

We compare the prediction intervals under the three settings we discuss above (perfect, unbiased, biased), with plug-in prediction intervals and bootstrap prediction intervals under each setting. We choose $\delta = 0.05$ (corresponding to 95\% prediction intervals) for both plug-in and bootstrap, and $\alpha = 0.05$ (corresponding to confidence level of 95\%) for bootstrap.

To compare the prediction performance of each proposed model with real data, we choose $r=7$ (on each Monday we make ACU, ICU predictions for the next Monday), comparing with the actual value. Also we set $n$ increased by $1$ for each additional observed day in each model. We set algorithm parameters $b_0=1000, m=300, \delta=0.05,\alpha=0.05$. With these parameters, the perfect model, unbiased model and biased model proposed converged in $1.25$, $4.08$ and $4.25$ seconds, respectively, when run on a laptop.

\begin{figure*}
  \centering
  \includegraphics[width=1\linewidth]{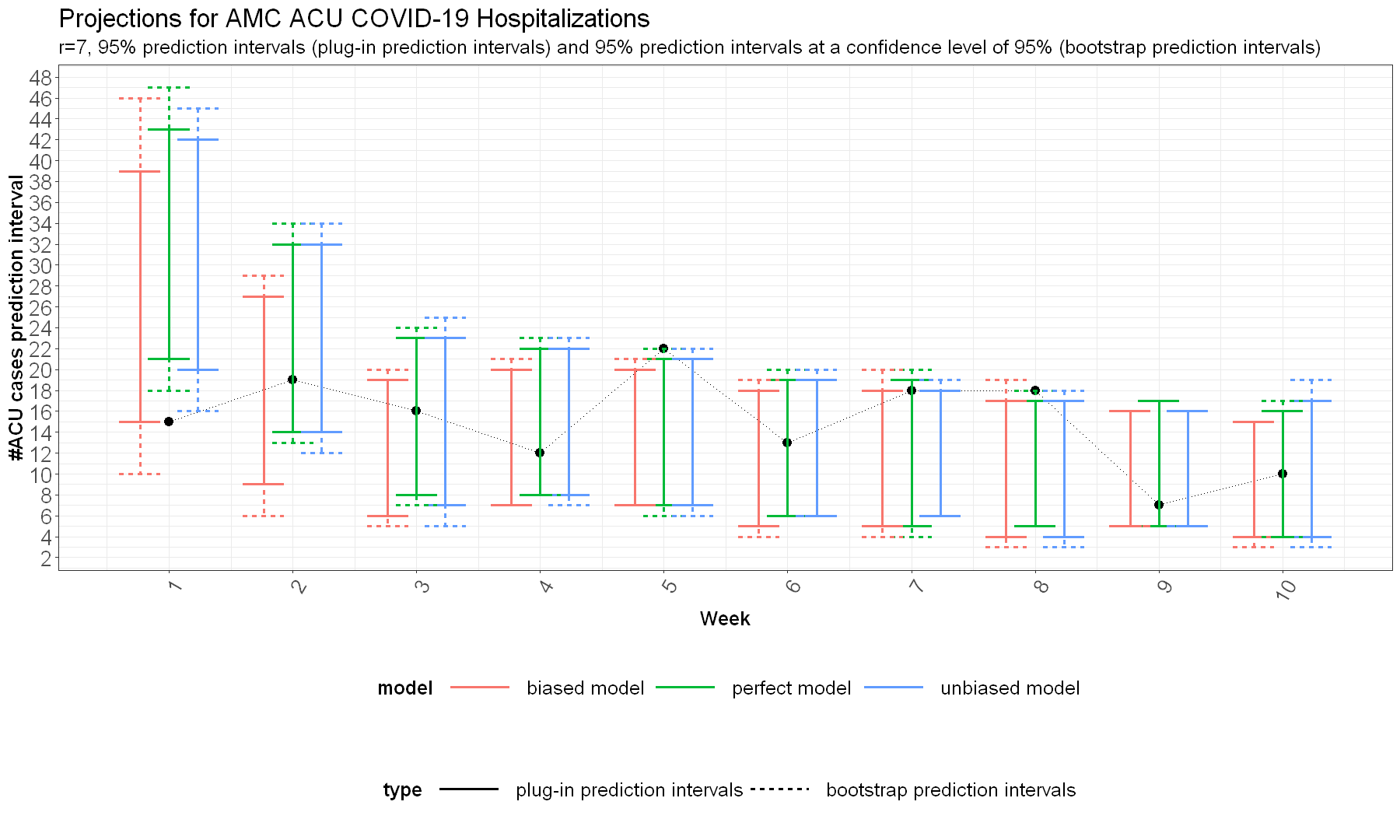}
  \caption{AMC ACU projections, $r=7$, 95\% prediction intervals, with black dots representing actual values}
  \label{fig:ACU-projections-r7}
\end{figure*}

\begin{figure*}
  \centering
  \includegraphics[width=1\linewidth]{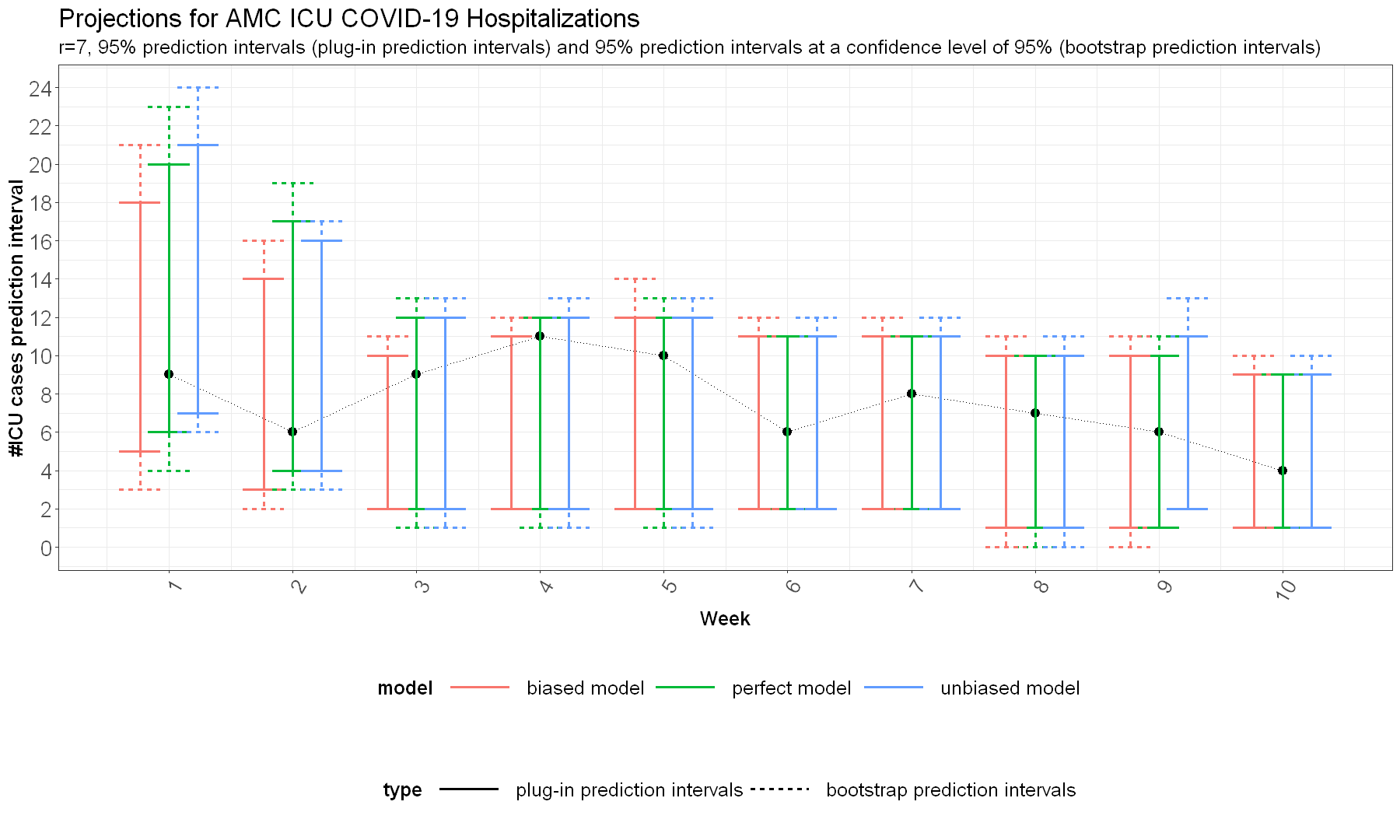}
  \caption{AMC ICU projections, $r=7$, 95\% prediction intervals, with black dots representing actual values}
  \label{fig:icu-projections-r7}
\end{figure*}

Projections for ACU and ICU made by different models are shown in Figure~\ref{fig:ACU-projections-r7} and~\ref{fig:icu-projections-r7}. In each plot, dash lines indicate $95\%$ bootstrap prediction intervals, solid lines indicate $95\%$ plug-in prediction intervals and black dots indicate actual values.
The unbiased models tend to provide wider prediction intervals. As we get larger $n$, the bootstrap prediction intervals are getting closer to the plugin prediction intervals.



\begin{figure*}
  \centering
  \includegraphics[width=1\linewidth]{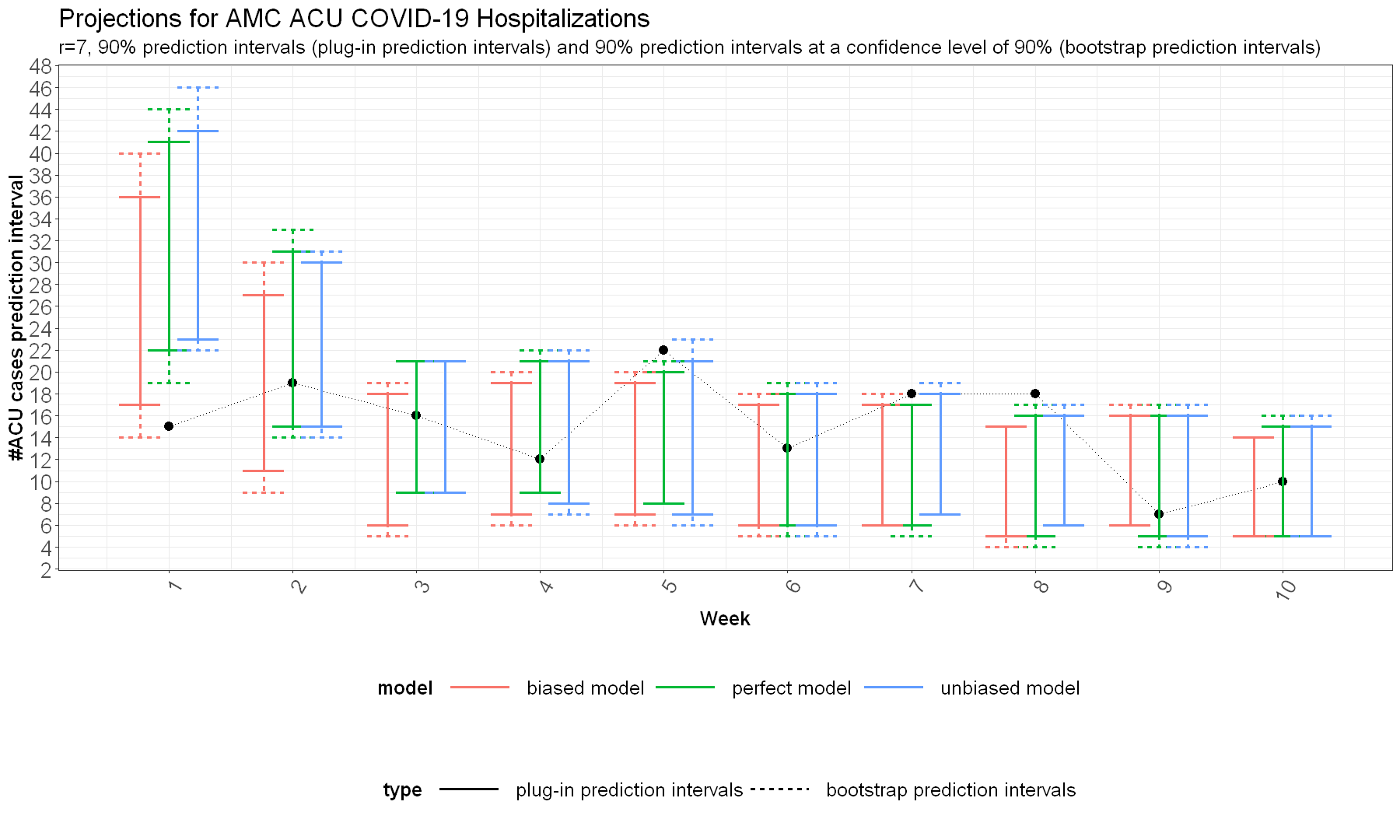}
  \caption{AMC ACU projections, $r=7$, 90\% prediction intervals, with black dots representing actual values}
  \label{fig:ACU-projections-r7-90}
\end{figure*}
\begin{figure*}
  \centering
  \includegraphics[width=1\linewidth]{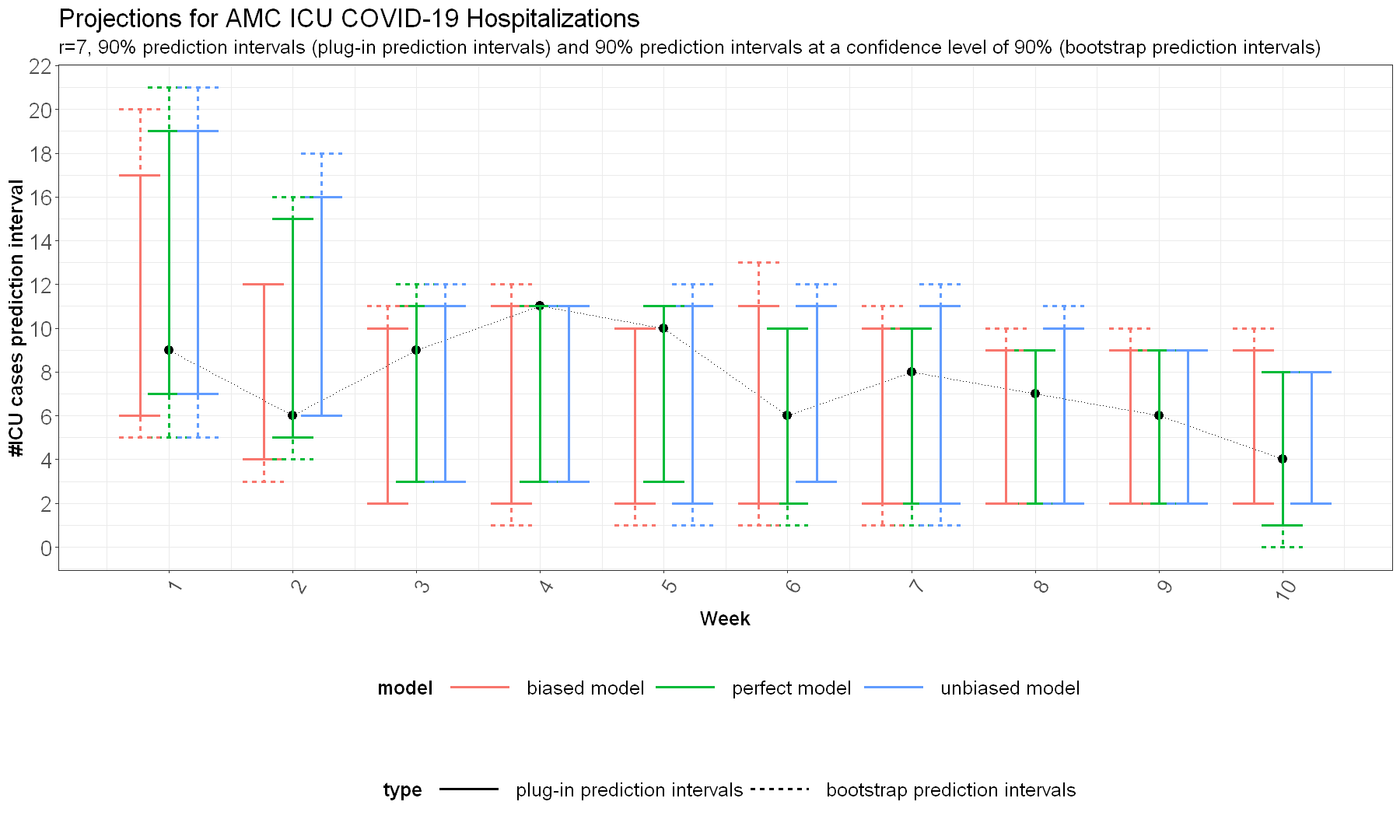}
  \caption{AMC ICU projections, $r=7$, 90\% prediction intervals, with black dots representing actual values}
  \label{fig:icu-projections-r7-90}
\end{figure*}
\begin{figure*}
  \centering
  \includegraphics[width=1\linewidth]{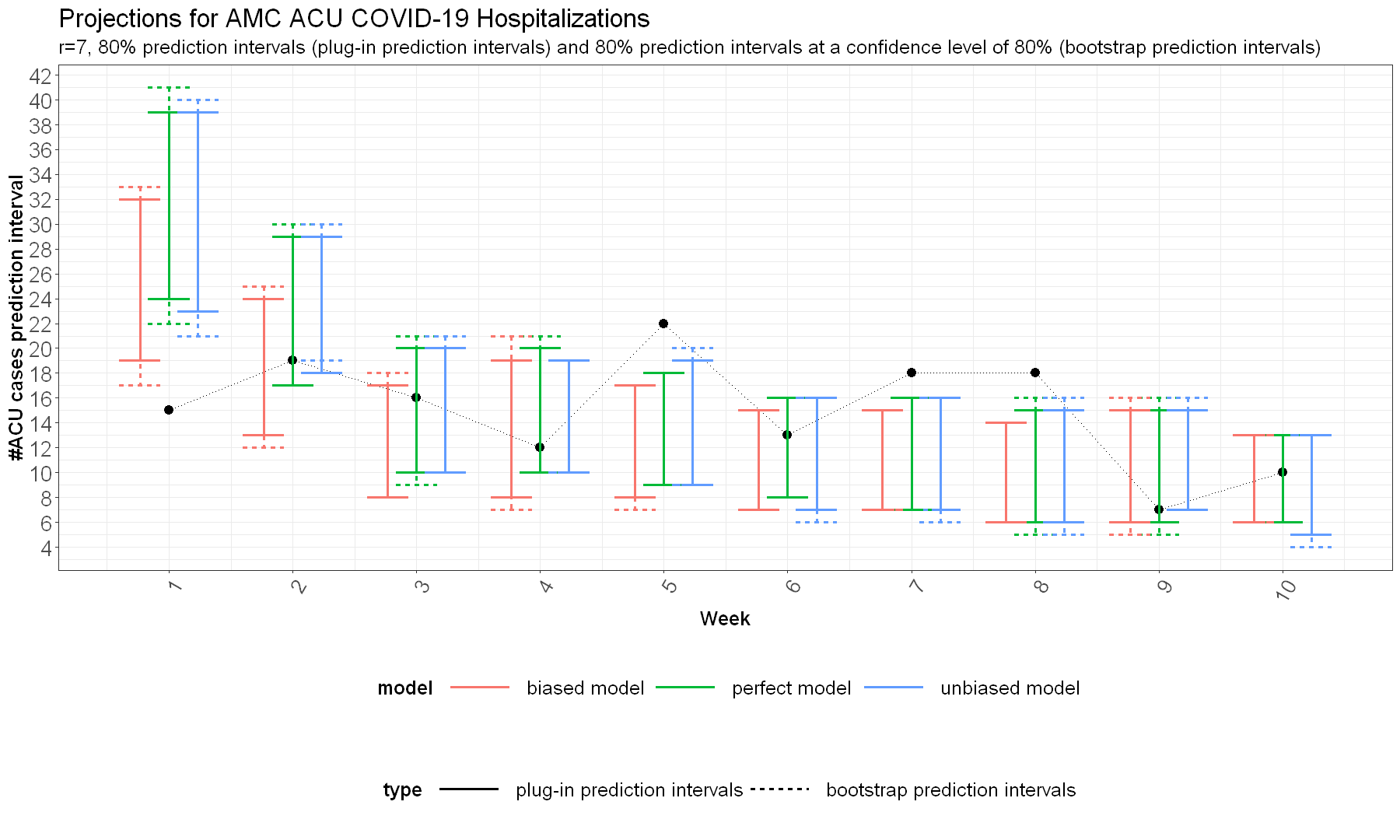}
  \caption{AMC ACU projections, $r=7$, 80\% prediction intervals, with black dots representing actual values}
  \label{fig:ACU-projections-r7-80}
\end{figure*}

As shown in Table ~\ref{tab:amc_95ci}, with $r=7$, the fractions of weeks for which each $95\%$ plug-in prediction intervals covered the observed bed count in the ACU is $70\%$ for all three models. The $95\%$ bootstrap prediction intervals covered $90\%$ of the observed bed count in the ACU for all models. All of the prediction intervals covered $100\%$ of the observed bed count in the ICU. Bootstrap intervals are generally wider, thus the coverage rates are higher. The results with $90\%$ prediction intervals (Figure~\ref{fig:ACU-projections-r7-90}, Figure~\ref{fig:icu-projections-r7-90}, Table~\ref{tab:amc_90ci}) and $80\%$ prediction intervals (Figure~\ref{fig:ACU-projections-r7-80}, Figure~\ref{fig:icu-projections-r7-80}, Table~\ref{tab:amc_80ci}) on AMC data can be found in the Appendix.

The demand intervals forecast using the perfect model were communicated to the hospital manager in charge of COVID-19 response capacity planning. The upper bound of the prediction intervals remained below the threshold hospital leadership felt comfortable could be accommodated without the cancellation of elective admissions.
\begin{table*}
\centering
\begin{tabular}{l|l|l|l|l}
\hline\noalign{\smallskip}
Model & Plug-in, ACU & Bootstrap, ACU& Plug-in, ICU & Bootstrap, ICU \\
\noalign{\smallskip}\hline \hline\noalign{\smallskip}
Perfect Model & $70\%$ & $90\%$ & $100\%$ & $100\%$ \\
Unbiased Model & $70\%$ & $90\%$ & $100\%$ & $100\%$\\
Biased Model & $70\%$ & $90\%$ & $100\%$ & $100\%$\\
\noalign{\smallskip}\hline
\end{tabular}
\caption{Coverage rate of 95\% plug-in prediction intervals and 95\% bootstrap prediction intervals at a confidence level of 95\%, AMC}
\label{tab:amc_95ci}       
\end{table*}



\subsection{Performance Evaluation: Synthetic Data}
In this section, we generate synthetic data for 100 days. The $\lambda$'s are generated using $SIR$ model (see, for example, \cite{bartlett1956deterministic} and \cite{allen1994some}), and the forecasts are generated following different model assumptions. We generate $N_i$ from the Poisson distribution with mean  $\lambda_i$. $A_i$'s and $B_i$'s are generated from the multinomial distribution with parameters $(N_i,p,q,1-p-q)$. They are all generated once and used in all following sections.

To evaluate the performances, we apply the above prediction methods on the last 60 observations.

\begin{figure*}
        \centering
        \includegraphics[width=1\textwidth]{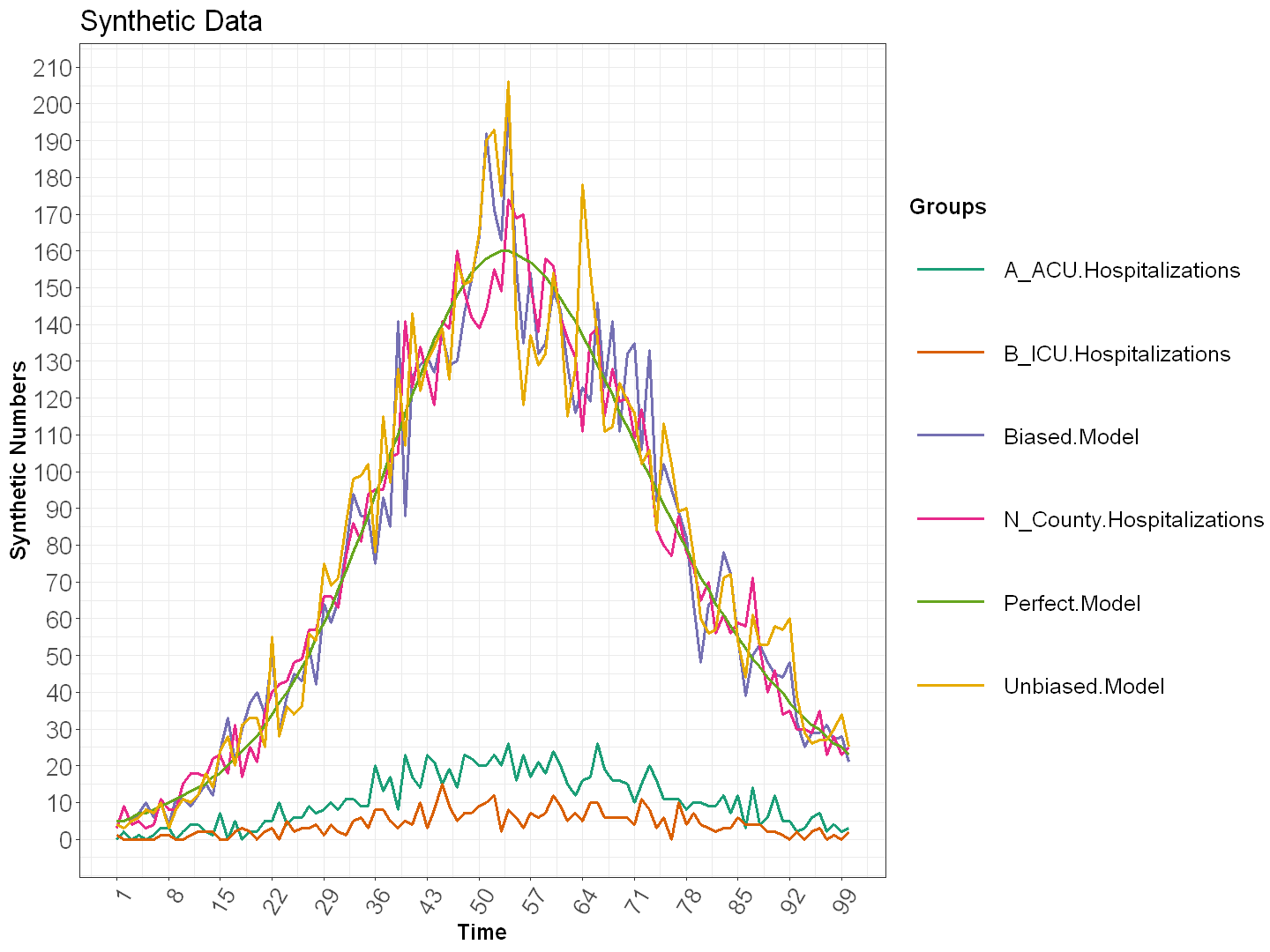}
        \caption{Synthetic Data}
  \label{fig:synthetic-data}
    \end{figure*}

The synthetic data are shown in Figure~\ref{fig:synthetic-data}.

\subsubsection{Synthetic Data under Perfect Forecasts Model}
\begin{figure*}
        \centering
        \includegraphics[width=1\textwidth]{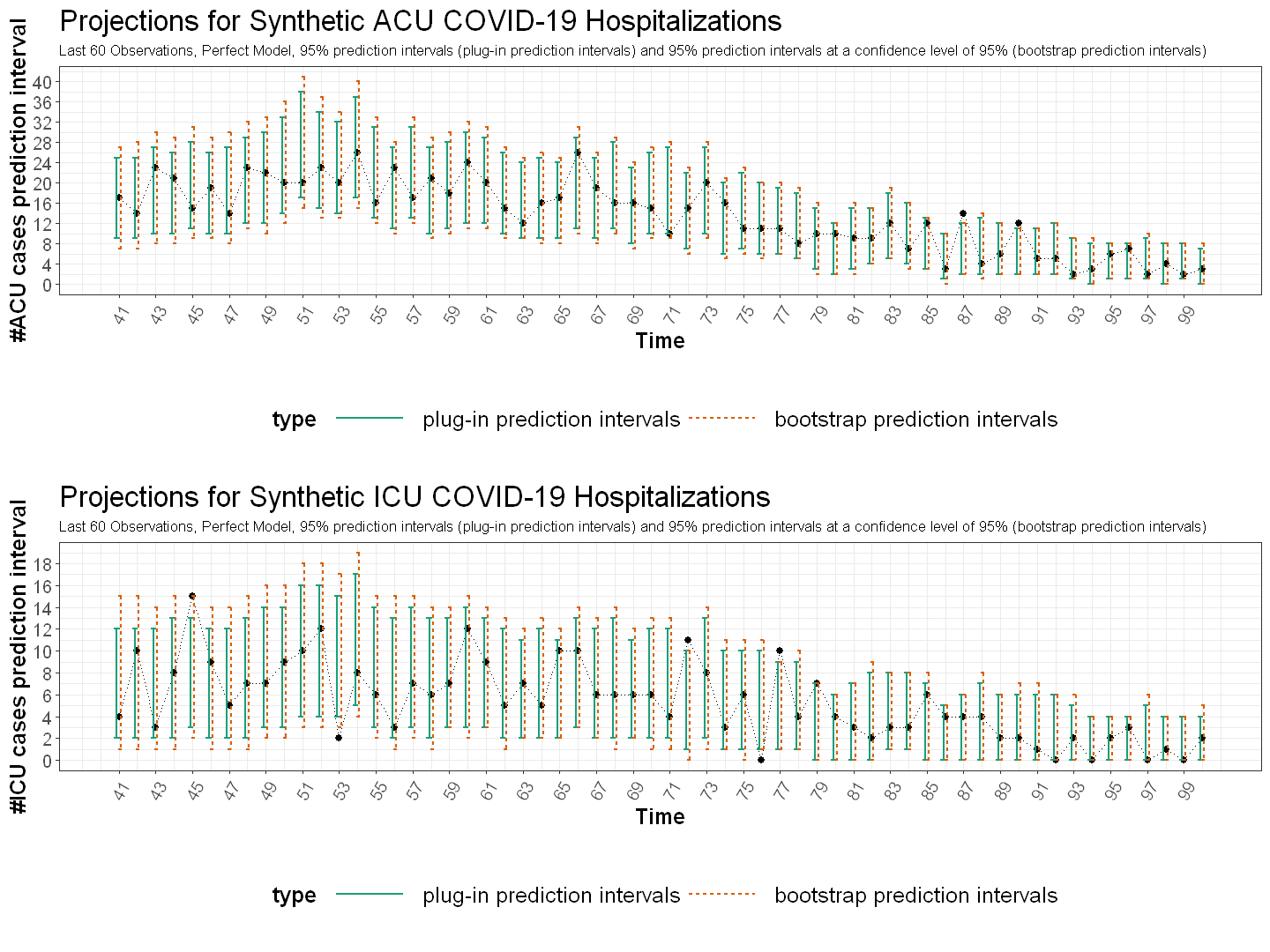}
        \caption{Projections on synthetic data, 95\% prediction intervals, perfect model, with black dots representing actual values}
        \label{fig:perfect-model-synthetic}
\end{figure*}

Here we generate
\begin{align*}
    F_i&=\lambda_i,  i =1,...,100
\end{align*} satisfying the ``perfect forecast'' assumption. The $95\%$ prediction intervals  for ACU($\{A_i\}$), ICU($\{B_i\}$) are shown in Figure~\ref{fig:perfect-model-synthetic}. The fractions of observations for which $95\%$ plug-in prediction intervals covered the observed bed count are $97\%$, $90\%$ for ACU, ICU respectively; the ones for $95\%$ bootstrap prediction intervals are $98\%$, $93\%$.

\subsubsection{Synthetic Data under Unbiased Forecasts Model}
\begin{figure*}
        \centering
        \includegraphics[width=1\textwidth]{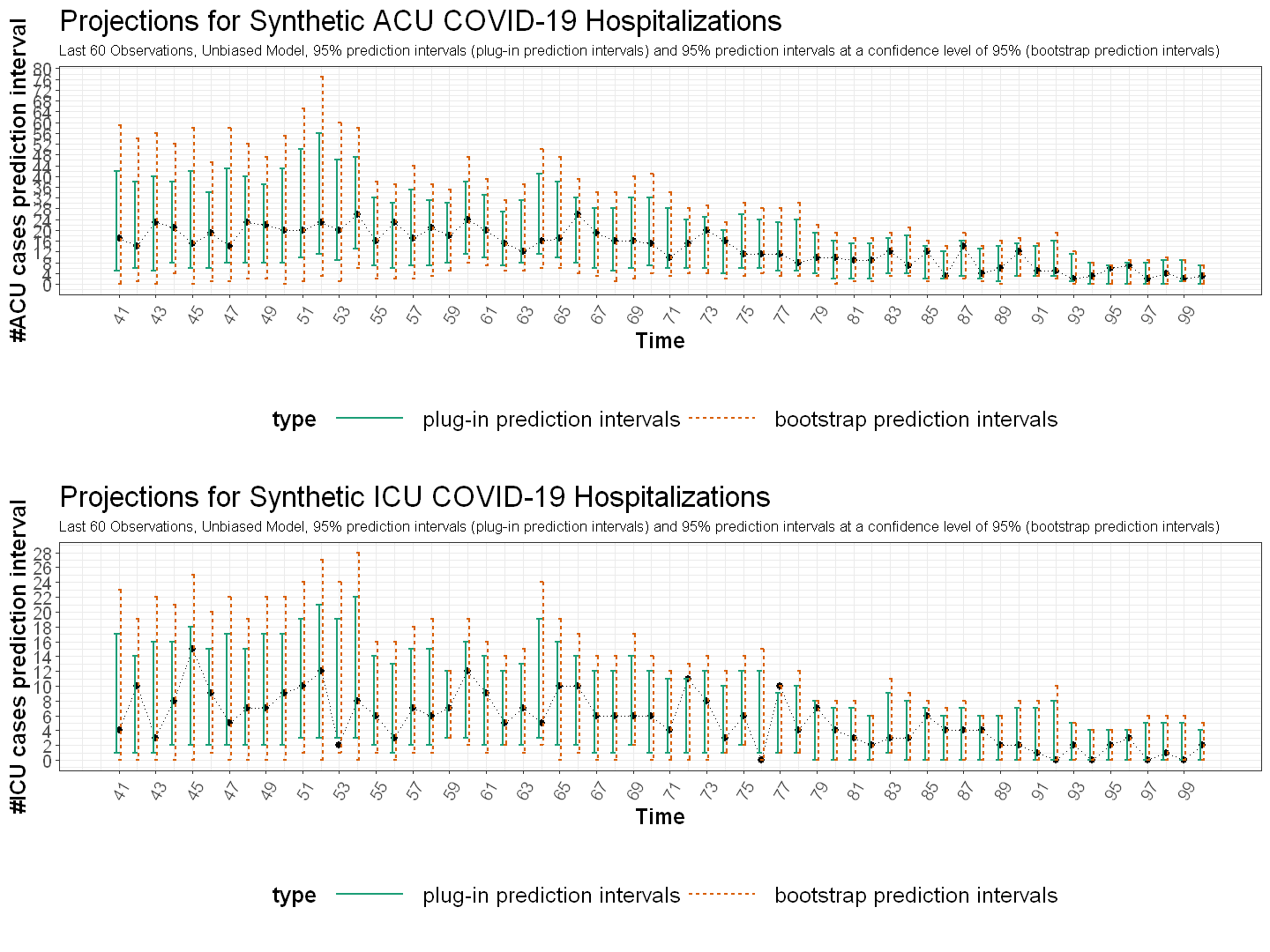}
        \caption{Projections on synthetic data, 95\% prediction intervals, unbiased model, with black dots representing actual values}
        \label{fig:unbiased-model-synthetic}
\end{figure*}

Under this setting, we set $\rho=0.5, \sigma^2=0.01,\mu=-\frac{\sigma^2}{2(1+\rho)}$ and generate
\begin{align*}
    Y_1 &\sim N(\frac{\mu}{1-\rho},\frac{\sigma^2}{1-\rho^2})\\
    Y_i&=\rho Y_{i-1}+Z_i, Z_i\sim N(\mu,\sigma^2), i =2,...,100\\
    \Gamma_i&=exp(Y_i),F_i=\frac{\lambda_i}{\Gamma_i},i =1,...,100,
\end{align*}
where $\sim$ represents ``distributed according to'', so that $\E(\Gamma_i)=1$ which satisfies the assumptions in the ``unbiased forecasts model''. The $95\%$ prediction intervals for ACU($\{A_i\}$), ICU($\{B_i\}$) are shown in Figure~\ref{fig:unbiased-model-synthetic}. The fractions of observations for which $95\%$ plug-in prediction intervals covered the observed bed count are $100\%$, $98\%$ for ACU, ICU respectively; the ones for $95\%$ bootstrap prediction intervals are $100\%$, $100\%$. Since unbiased model tend to generate wider prediction intervals, the coverage rates are higher.

\subsubsection{Synthetic Data under Biased Forecasts Model}
\begin{figure*}
        \centering
        \includegraphics[width=1\textwidth]{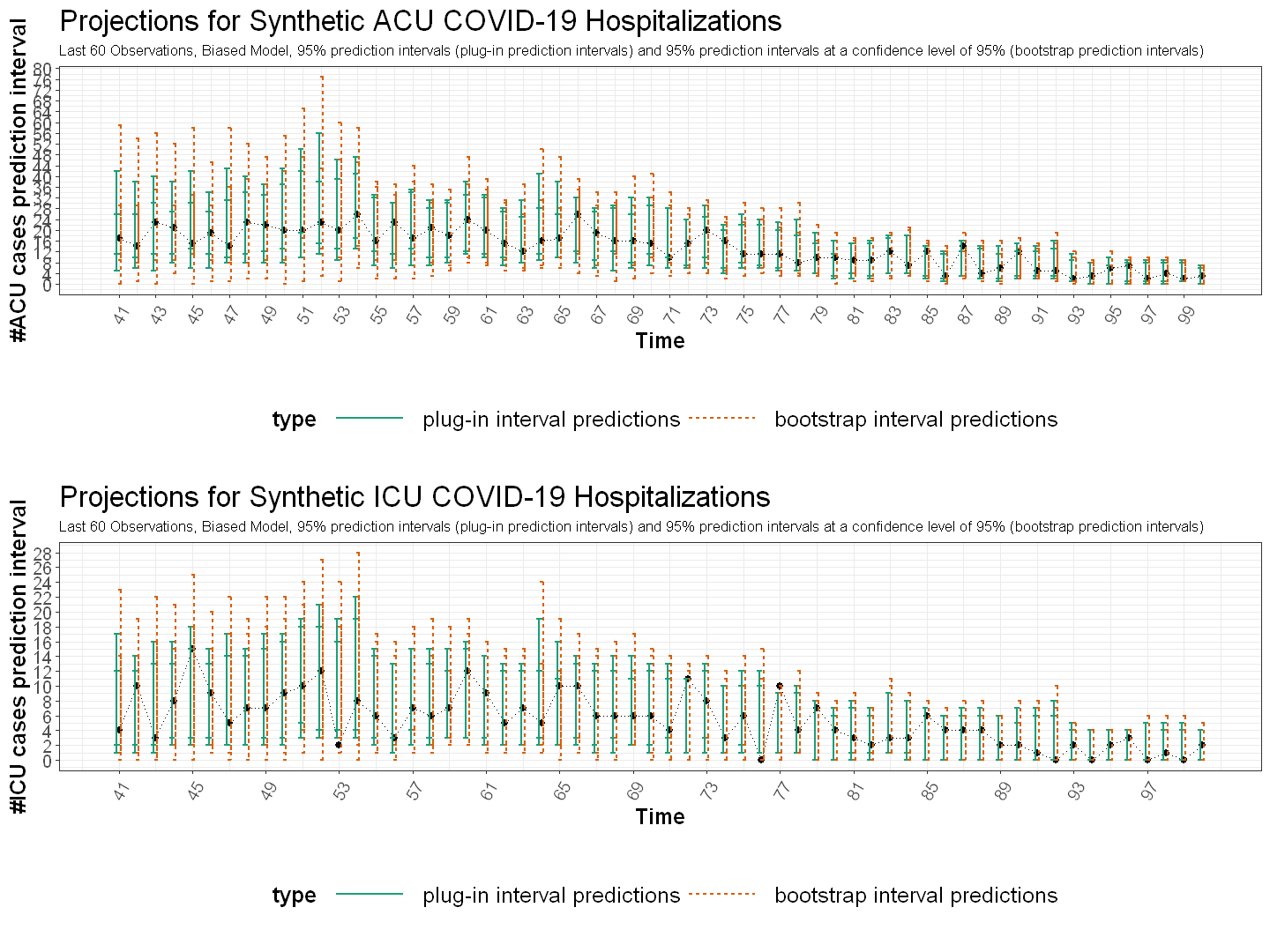}
        \caption{Projections on synthetic data, 95\% prediction intervals, biased model, with black dots representing actual values}
        \label{fig:biased-model-synthetic}
    \end{figure*}

Under this setting, we set $\rho=0.5, \sigma^2=0.01,\mu=0$. The generation method is the same as that in unbiased model setting. The $95\%$ prediction intervals for ACU($\{A_i\}$), ICU($\{B_i\}$) are shown in Figure~\ref{fig:biased-model-synthetic}. The fractions of observations for which plug-in prediction intervals covered the observed bed count are $97\%$, $97\%$ for ACU, ICU respectively; the ones for bootstrap prediction intervals are $100\%$, $98\%$.

\begin{figure*}
        \centering
        \includegraphics[width=0.8\textwidth]{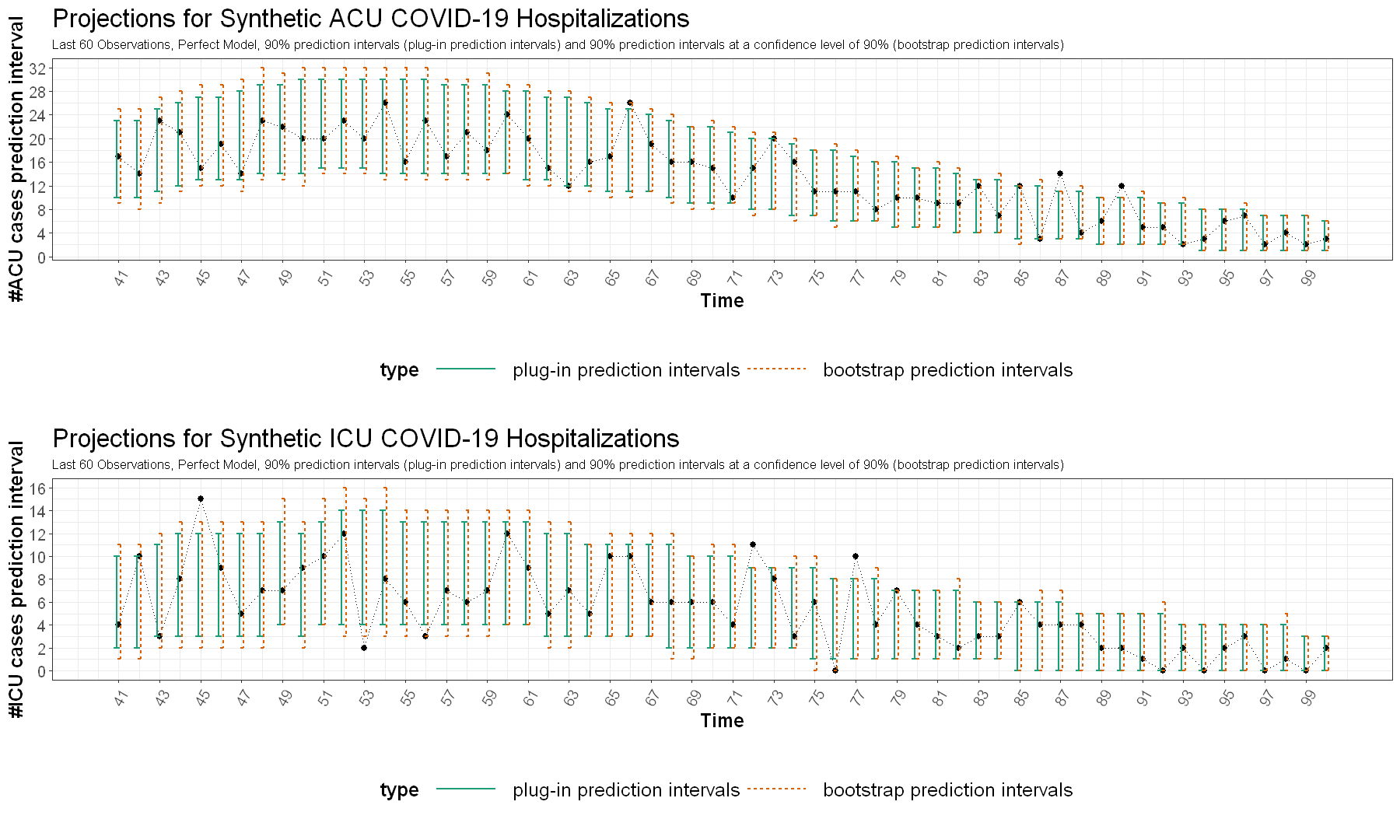}
        \caption{Projections on synthetic data, 90\% prediction intervals, perfect model, with black dots representing actual values}
        \label{fig:perfect-model-synthetic-90}
    \end{figure*}

\begin{figure*}
        \centering
        \includegraphics[width=1\textwidth]{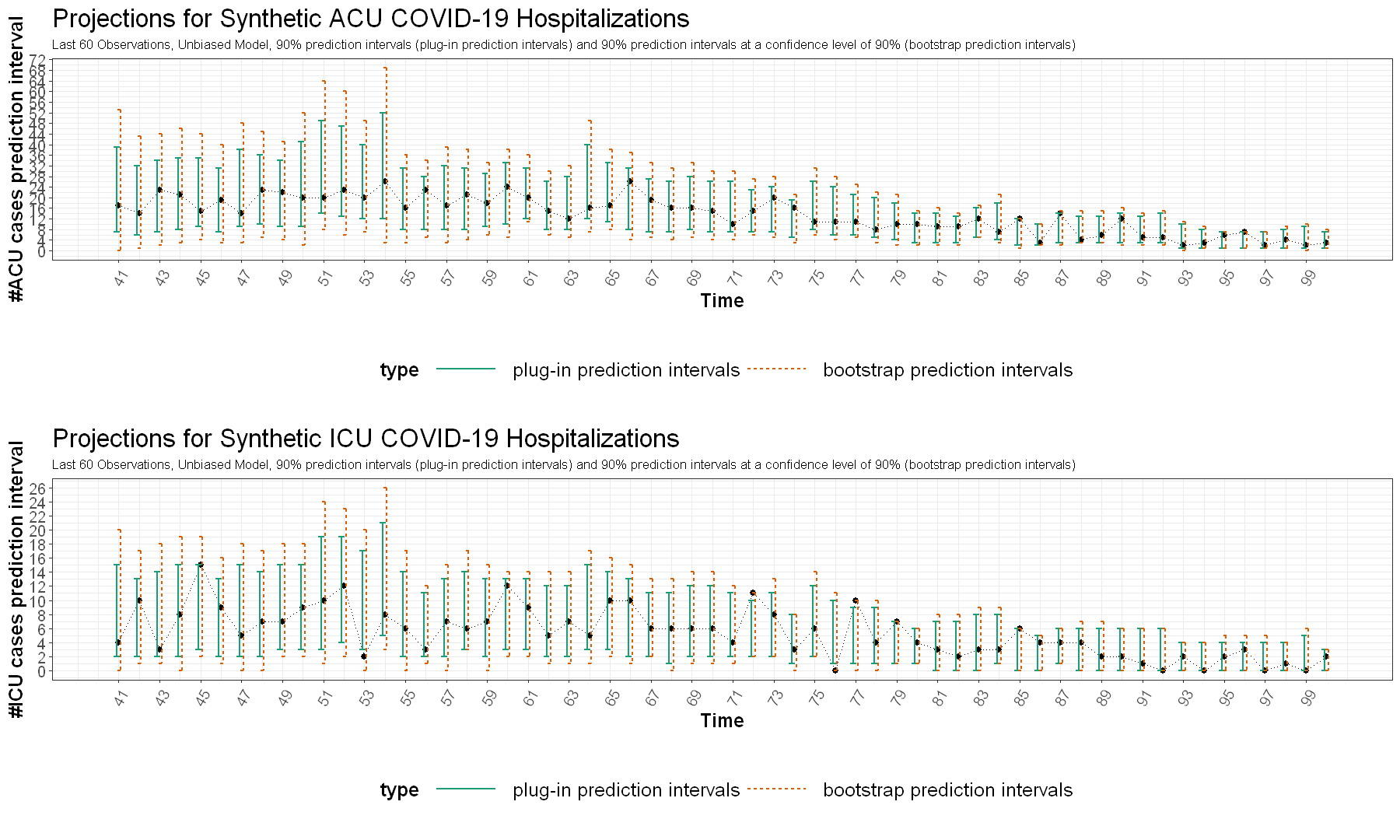}
        \caption{Projections on synthetic data, 90\% prediction intervals, biased model, with black dots representing actual values}
        \label{fig:unbiased-model-synthetic-90}
    \end{figure*}

\begin{figure*}
        \centering
        \includegraphics[width=1\textwidth]{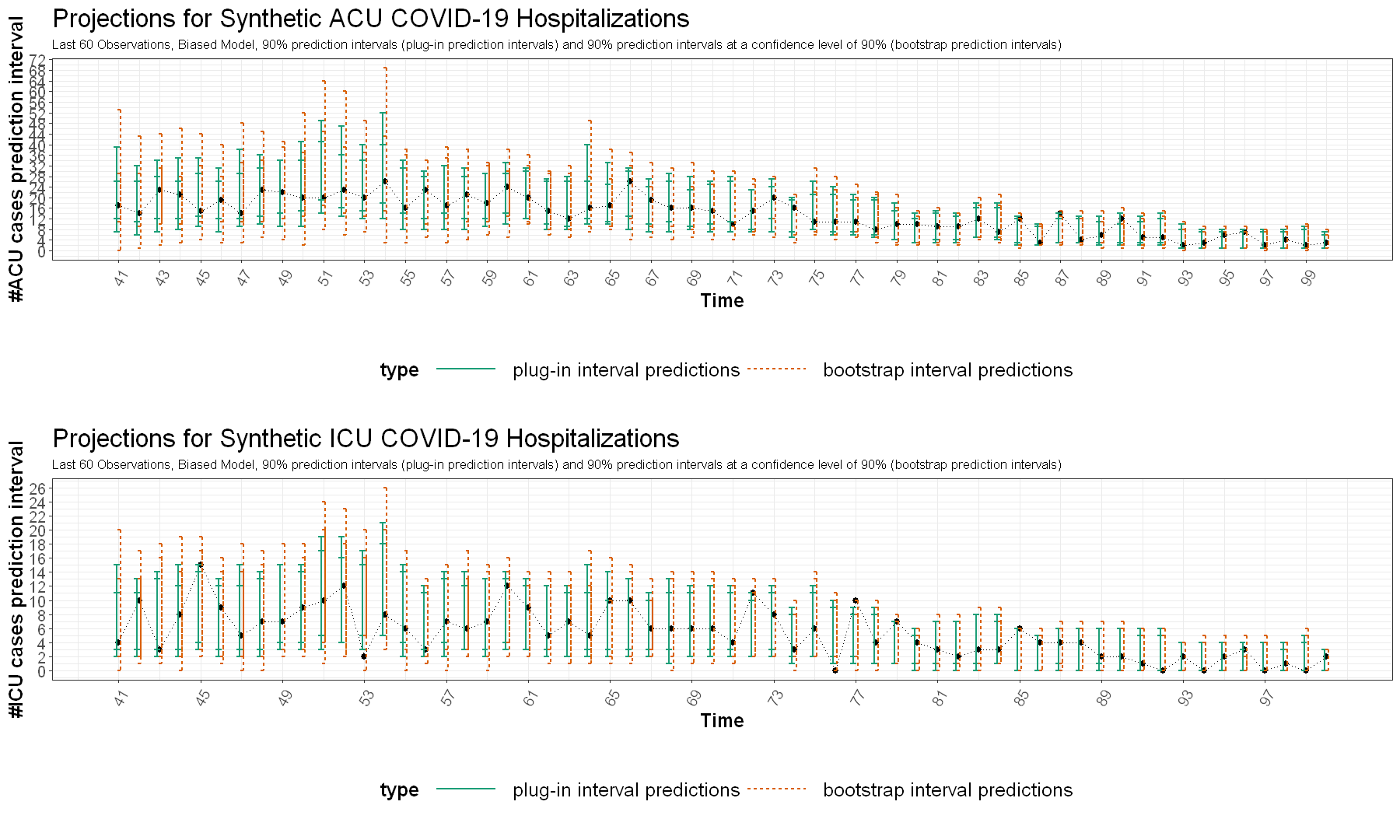}
        \caption{Projections on synthetic data, 90\% prediction intervals, biased model, with black dots representing actual values}
        \label{fig:biased-model-synthetic-90}
    \end{figure*}

\begin{figure*}
        \centering
        \includegraphics[width=1\textwidth]{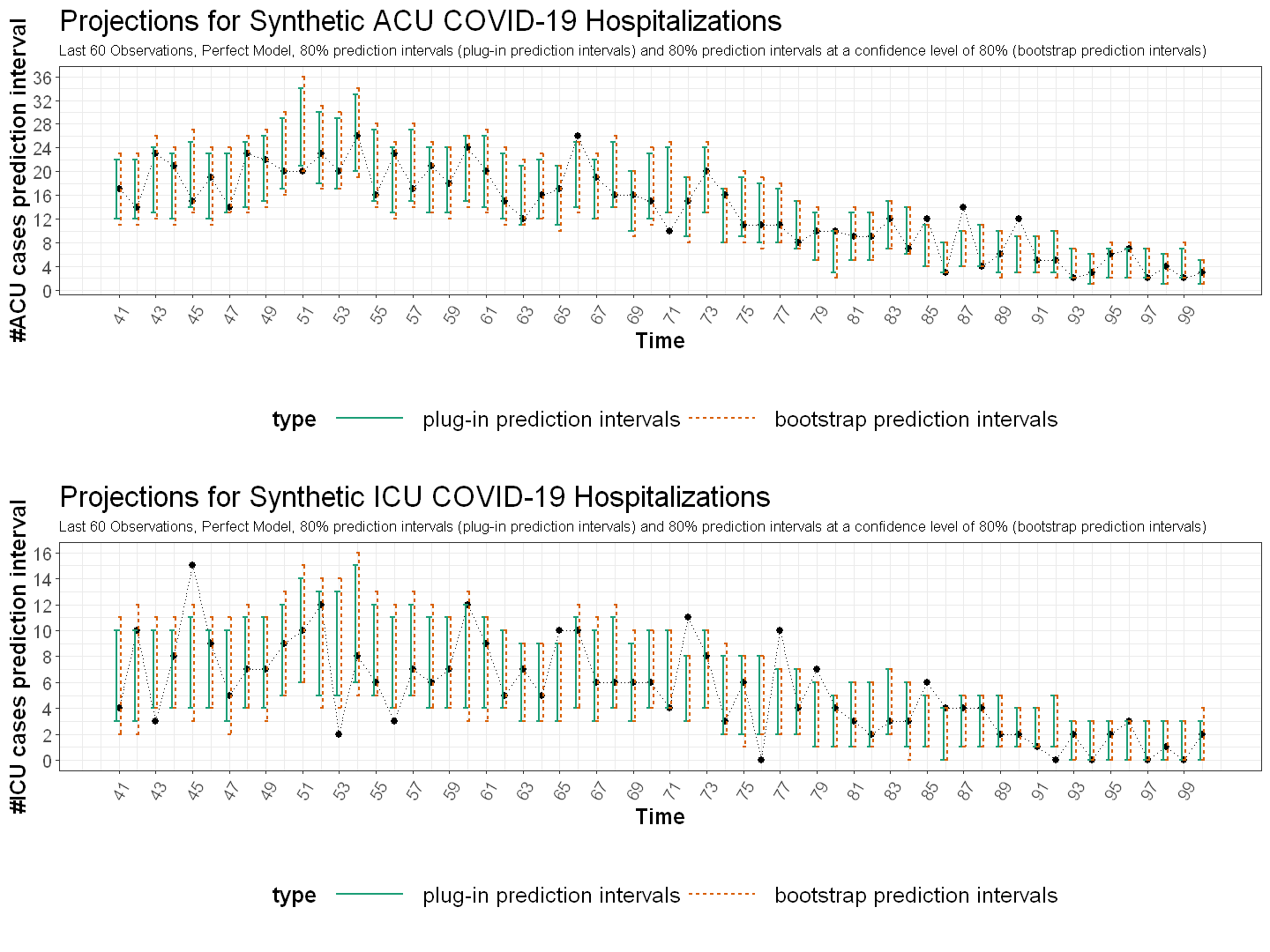}
        \caption{Projections on synthetic data, 80\% prediction intervals, perfect model, with black dots representing actual values}
        \label{fig:perfect-model-synthetic-80}
    \end{figure*}

\begin{figure*}
        \centering
        \includegraphics[width=1\textwidth]{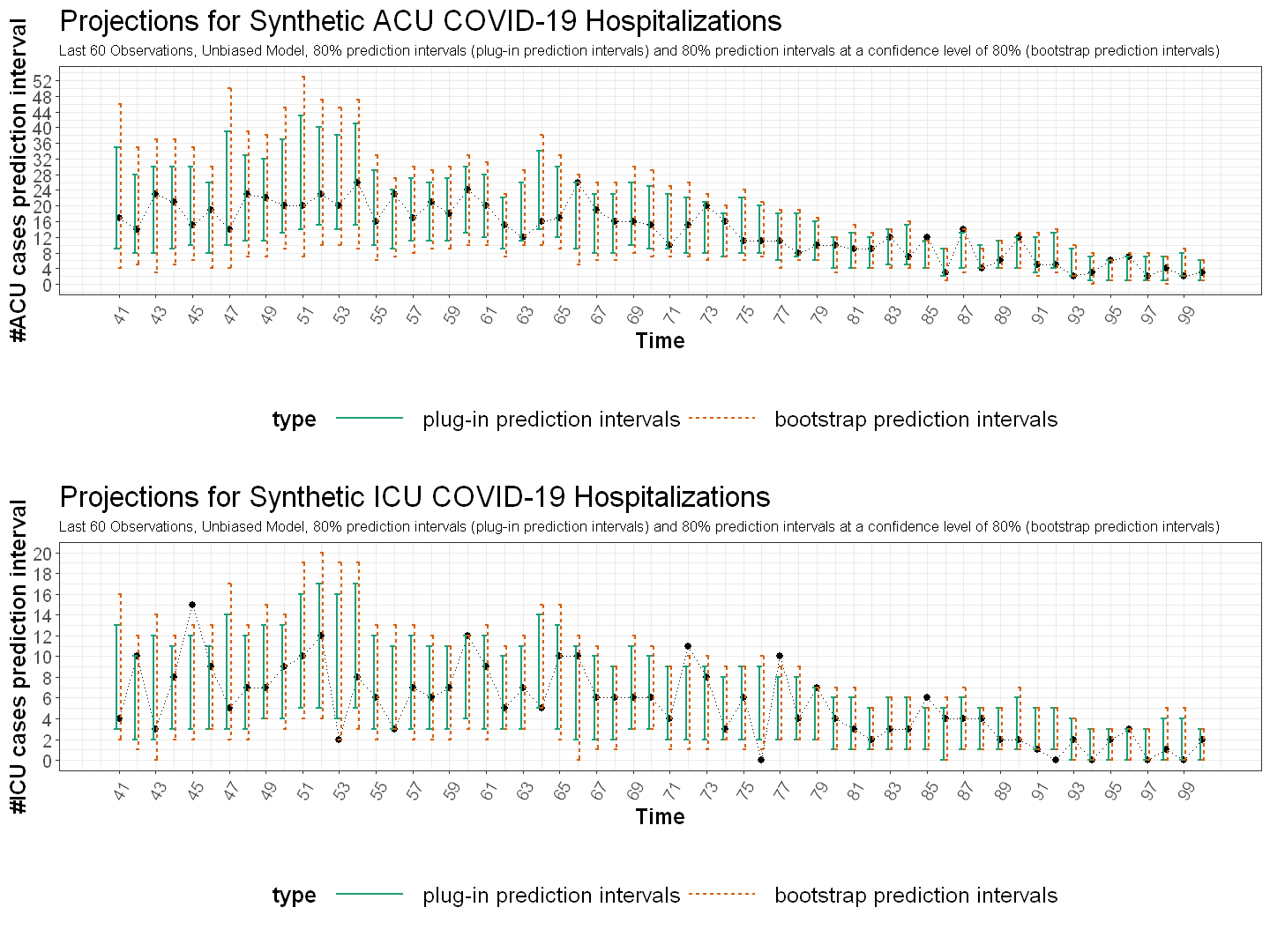}
        \caption{Projections on synthetic data, 80\% prediction intervals, unbiased model, with black dots representing actual values}
        \label{fig:unbiased-model-synthetic-80}
    \end{figure*}

All the plots show that with $n$ increasing, the bootstrap prediction intervals are getting closer to the plugin intervals. Unbiased and biased models larger variances, resulting in wider prediction intervals. The coverage rates of prediction intervals are shown in Table~\ref{tab:synthetic-95ci}. The results with $90\%$ prediction intervals (Figure~\ref{fig:perfect-model-synthetic-90}, Figure~\ref{fig:unbiased-model-synthetic-90},
Figure~\ref{fig:biased-model-synthetic-90}, Table~\ref{tab:synthetic-90ci}) and $80\%$ prediction intervals (Figure~\ref{fig:perfect-model-synthetic-80}, Figure~\ref{fig:unbiased-model-synthetic-80},
Figure~\ref{fig:biased-model-synthetic-80}, Table~\ref{tab:synthetic-80ci}) on synthetic data can be found in the Appendix.

\begin{table*}
\centering
\begin{tabular}{l|l|l|l|l}
\hline\noalign{\smallskip}
Model & Plug-in, ACU & Bootstrap, ACU& Plug-in, ICU & Bootstrap, ICU \\
\noalign{\smallskip}\hline\hline\noalign{\smallskip}
Perfect Model & $97\%$ & $98\%$ & $92\%$ & $95\%$ \\
Unbiased Model & $100\%$ & $100\%$ & $98\%$ & $100\%$\\
Biased Model & $97\%$ & $100\%$ & $97\%$ & $98\%$\\
\noalign{\smallskip}\hline
\end{tabular}
\caption{Coverage rate of 95\% prediction intervals, synthetic data}
\label{tab:synthetic-95ci}       
\end{table*}

\section{Conclusions}

In this work we introduce, DICE (Demand Intervals from Consistent Estimators), a model to forecast prediction intervals for the fraction of regional patient demand arriving to an institution based on the historical fraction of demand served by the institution and, potentially biased, forecasts of demand as a Poisson random variable. We show that our model is consistent, computationally tractable, and well-calibrated on empirical data as well as synthetic data. Unlike Hidden Markov models that may fail to converge or be prohibitively computationally expensive due to the curse of dimensionality, all of the models tested converged in seconds. To illustrate its potential usefulness, we discuss the managerial COVID-19 decisions that prompted the development of the models as well as how they were used to inform these decisions at an academic medical center. The demand interval forecasts suggested that the `second wave' influx of COVID-19 patients would be unlikely to exceed available hospital capacity. The information provided by the model contributed to, the ultimately correct, decision that COVID-19 patients could be accommodated without the cancellation of elective admissions.

As hospitals the world over prepare for a third wave of COVID-19, this model may find similar applications at institutions planning their response to an influx of patients. Beyond COVID-19, patient demand for a variety of medical conditions is forecast as a Poisson random variable. The model developed may be of use to the numerous decision to make which hospital managers project demand for their services by combining their historical share of regional demand with forecasts of total regional demand.

\medskip
\bibliographystyle{spbasic_unsorted}
\bibliography{notebib}

\newpage

\section*{Appendix}

We establish here that the ``method of moments'' estimators of Sections 4 and 5 will be consistent in great generality. This will follow if we can prove that
\begin{align}\label{eq:append-m_i-converge-in-prob}
    \hat M_i
    \xrightarrow{p}
    m_i(\mu_0, \sigma_0^2, \rho_0) \tag{A.1}
\end{align}
as $n \to \infty$, for $1 \le i \le 3$, where $\xrightarrow{p}$ denotes ``converge in probability''.

Note that $\E \hat M_i = m_i(\mu_0, \sigma_0^2, \rho_0)$ for $1 \le i \le 3$.
Hence, \eqref{eq:append-m_i-converge-in-prob} follows from Chebyshev's inequality if we can show that
\begin{equation*}
    \begin{aligned}
    \Var \hat M_i \to 0
    \end{aligned}
\end{equation*}
as $n \to \infty$. But
\begin{align*}
     \Var \hat M_1
     & = \frac{1}{n^2} \Var \left( \sum_{i = -n}^{-1} \frac{N_i}{F_i}  \right) \\
     & =\frac{1}{n^2} \sum_{i = -n}^{-1} \Var \left( \frac{N_i}{F_i}  \right)
     + \\
     & \ \ \ \  \frac{2}{n^2} \sum_{i = -n}^{-1}\sum_{j = i+1}^{-n} \Cov \left(
     \frac{N_i}{F_i}, \frac{N_j}{F_j}
     \right)
\end{align*}
Of course,
\begin{align*}
    \Var \left(
     \frac{N_i}{F_i}
     \right)
     & = \E \left(\frac{N_i^2}{F_i^2} \right) -
      \left ( \E \left(\frac{N_i}{F_i} \right) \right) ^2\\
     & = \frac{\E N_i^2}{ \lambda_i^2} \cdot \E \Gamma_i^2 - m_1(\mu_0, \sigma_0^2, \rho_0)^2 \\
     & = \frac{\lambda_i + \lambda_i^2}{ \lambda_i^2} \cdot \E \Gamma_{-1}^2  \\
     & = \Var \Gamma_{-1} + \frac{1}{\lambda_i} \E \Gamma_{-1}^2.
\end{align*}
Also, for $i < j$
\begin{align*}
    \Cov \left(
     \frac{N_i}{F_i}, \frac{N_j}{F_j}
     \right)
     & = \E \left(
     \frac{N_i}{F_i} \frac{N_j}{F_j}
     \right)  -
     \E \left(
     \frac{N_i}{F_i}\right) \E \left(  \frac{N_j}{F_j}
     \right) \\
     & = \frac{\E N_i \E N_j}{\lambda_i \lambda_j} \E \Gamma_i \Gamma_j -
     m_1(\mu_0, \sigma_0^2, \rho_0)^2 \\
     & = \Cov \left(
     \Gamma_{-1 - (j-i)}, \Gamma_{-1} \right) \\
     & = \E \exp ( Y_{-1 - (j-i)} + Y_{-1} ) - \left(\E \Gamma_{-1} \right)^2 \\
     & = \E \exp (
     (1 + \rho^{-j-i}) Y_{-1 - (j-i)} + \\
     & \ \ \ \  \sum_{j = 0}^{j-i-1} \rho^j Z_{-1-j}
     )
     - \left(\E \Gamma_{-1} \right)^2 \\
     & = O(\rho^{j-i}),
\end{align*}
where $O(a_i)$ denotes a quantity that is bounded by a multiple of $|a_i|$. Similar calculation can be found in, for example, \cite{hamilton1994time}. Consequently,
\begin{align*}
    \Var \hat M_1
    & = \frac{1}{n^2}
    \sum_{i = -n}^{-1} \left[
    \Var \Gamma_{-1} +\frac{1}{\lambda_i} \E \Gamma_{-1}^2
    \right]
    + \\
    & \ \ \ \ \frac{2}{n^2} \sum_{i = -n}^{-1}\sum_{j = i+1}^{-n} O(\rho^{j-i}) \\
    & = O\left(\frac{1}{n} \right) \to 0
\end{align*}
as $n \to \infty$ if we assume that the infimum of the $\lambda_i$'s is bounded away from zero.
Similarly, $\Var \hat M_i \to 0$ for $i = 2$ and $i = 3$ under this very moderate hypothesis on the $\lambda_i$'s, thereby establishing the consistency.

\begin{table*}
\centering
\begin{tabular}{l|l|l|l|l}
\hline\noalign{\smallskip}
Model & Plug-in, ACU & Bootstrap, ACU& Plug-in, ICU & Bootstrap, ICU \\
\noalign{\smallskip}\hline\hline\noalign{\smallskip}
Perfect Model & $60\%$ & $60\%$ & $100\%$ & $100\%$ \\
Unbiased Model & $60\%$ & $60\%$ & $100\%$ & $100\%$\\
Biased Model & $70\%$ & $90\%$ & $100\%$ & $100\%$\\
\noalign{\smallskip}\hline
\end{tabular}
\caption{Coverage rate of 90\% prediction intervals, AMC}
\label{tab:amc_90ci}       
\end{table*}


\begin{table*}
\centering
\begin{tabular}{l|l|l|l|l}
\hline\noalign{\smallskip}
Model & Plug-in, ACU & Bootstrap, ACU& Plug-in, ICU & Bootstrap, ICU \\
\noalign{\smallskip}\hline\hline\noalign{\smallskip}
Perfect Model & $60\%$ & $60\%$ & $90\%$ & $90\%$ \\
Unbiased Model & $60\%$ & $70\%$ & $100\%$ & $100\%$\\
Biased Model & $60\%$ & $70\%$ & $90\%$ & $90\%$\\
\noalign{\smallskip}\hline
\end{tabular}
\caption{Coverage rate of 80\% prediction intervals, AMC}
\label{tab:amc_80ci}       
\end{table*}

\begin{figure*}
  \centering
  \includegraphics[width=1\linewidth]{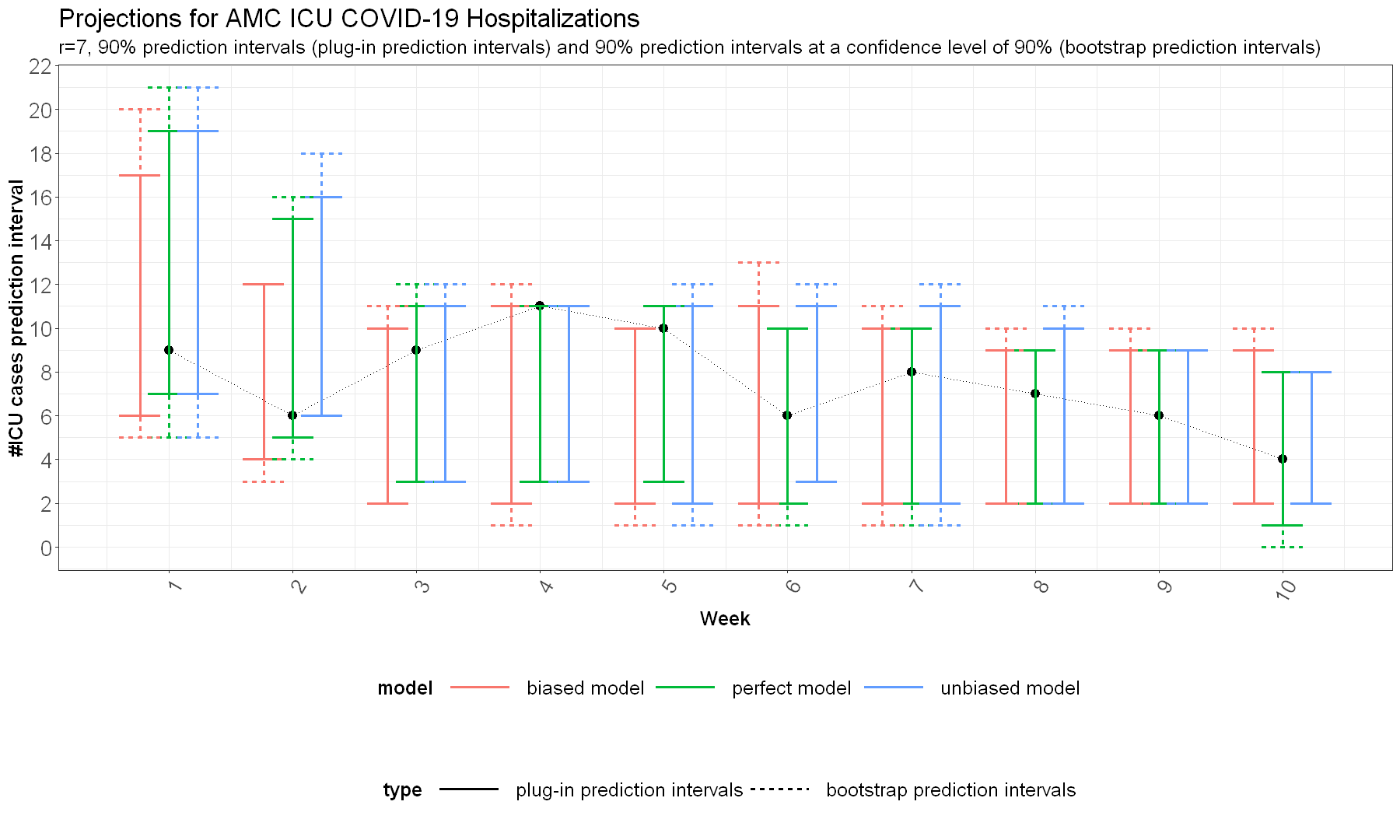}
  \caption{AMC ICU projections, $r=7$, 80\% prediction intervals, with black dots representing actual values}
  \label{fig:icu-projections-r7-80}
\end{figure*}

\begin{table*}
\centering
\begin{tabular}{l|l|l|l|l}
\hline\noalign{\smallskip}
Model & Plug-in, ACU & Bootstrap, ACU& Plug-in, ICU & Bootstrap, ICU \\
\noalign{\smallskip}\hline\hline\noalign{\smallskip}
Perfect Model & $95\%$ & $97\%$ & $90\%$ & $93\%$ \\
Unbiased Model & $98\%$ & $98\%$ & $92\%$ & $97\%$\\
Biased Model & $95\%$ & $97\%$ & $93\%$ & $95\%$\\
\noalign{\smallskip}\hline
\end{tabular}
\caption{Coverage rate of 90\% prediction intervals, synthetic data}
\label{tab:synthetic-90ci}       
\end{table*}

    \begin{figure*}
        \centering
        \includegraphics[width=1\textwidth]{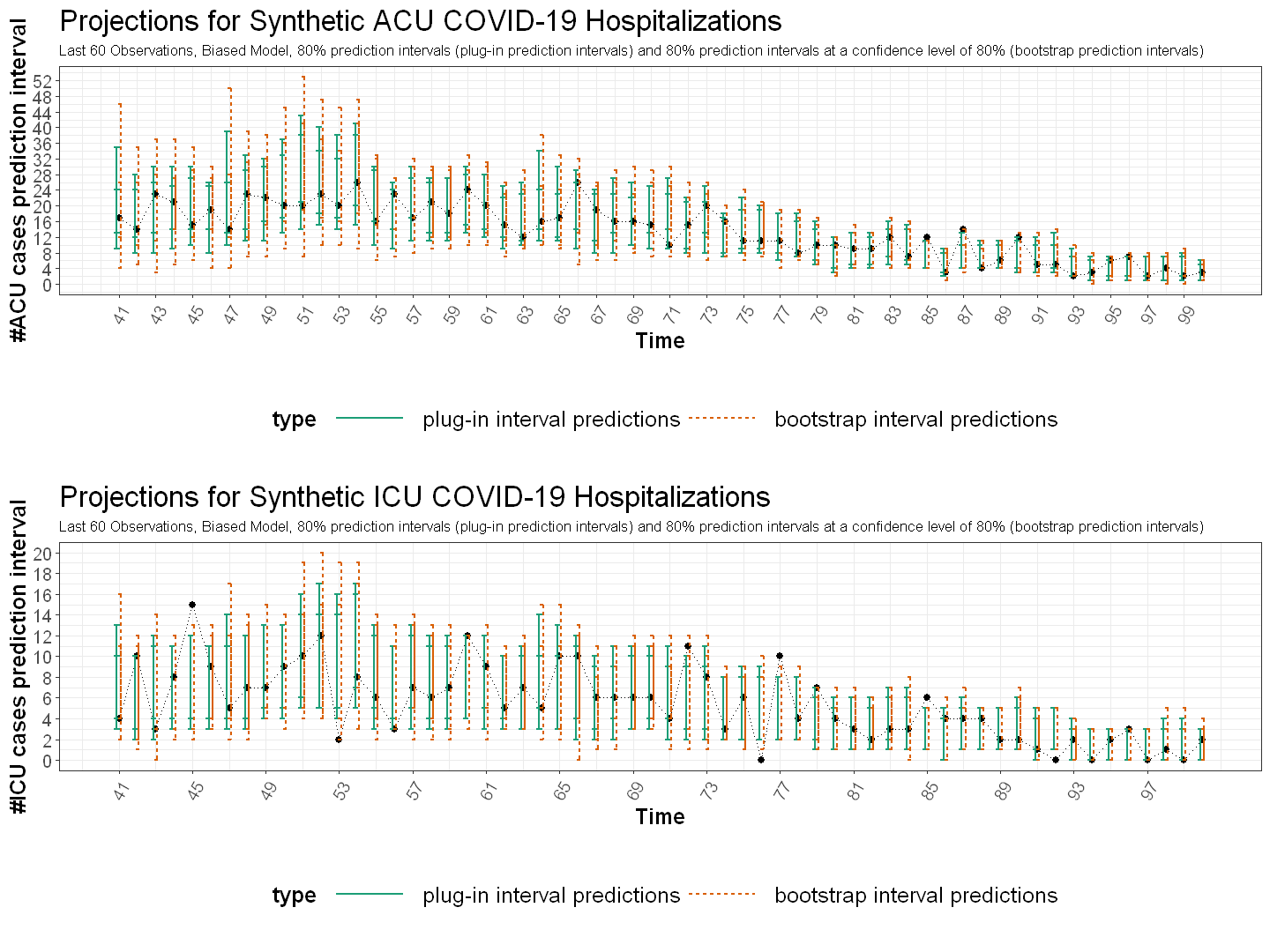}
        \caption{Projections on synthetic data, 80\% prediction intervals, biased model, with black dots representing actual values}
        \label{fig:biased-model-synthetic-80}
    \end{figure*}

\begin{table*}
\centering
\begin{tabular}{l|l|l|l|l}
\hline\noalign{\smallskip}
Model & Plug-in, ACU & Bootstrap, ACU& Plug-in, ICU & Bootstrap, ICU \\
\noalign{\smallskip}\hline\hline\noalign{\smallskip}
Perfect Model & $87\%$ & $88\%$ & $85\%$ & $87\%$ \\
Unbiased Model & $97\%$ & $97\%$ & $88\%$ & $90\%$\\
Biased Model & $92\%$ & $93\%$ & $82\%$ & $85\%$\\
\noalign{\smallskip}\hline
\end{tabular}
\caption{Coverage rate of 80\% prediction intervals, synthetic data}
\label{tab:synthetic-80ci}       
\end{table*}

%
%

\begin{acknowledgements}
We thank Kristan Lea Staudenmayer and jonathan Lu for help understanding the problem from the point of view of the hospital, accessing hospital data, and providing feedback about how the projections should be made useful.
\end{acknowledgements}

%
%



\end{document}